\documentclass[prb,twocolumn,longbibliography,superscriptaddress]{revtex4-2}
\pdfoutput=1
\usepackage{graphicx}
\usepackage{bm}
\usepackage{bbm}
\usepackage{wrapfig}
\usepackage{color}
\usepackage{natbib}
\usepackage{hyperref}
\usepackage{amssymb}
\usepackage{amsmath}
\usepackage{geometry}
\usepackage{comment}

\usepackage[capitalize]{cleveref} % more convenient citations
\usepackage{braket} % braket notation
\usepackage{mathtools} % paired delimiters
\usepackage{siunitx}

\DeclareMathOperator{\tr}{tr}
\DeclareMathOperator{\sgn}{sgn}
\DeclarePairedDelimiter\abs{\lvert}{\rvert}

\begin{document}
\title{Spin-valley collective modes of the electron liquid in graphene}

\author{Zachary M. Raines}
\affiliation{Department of Physics, Yale University, New Haven, CT 06520, USA}
\author{Vladimir I. Fal'ko}
\affiliation{National Graphene Institute, University of Manchester, Manchester M13 9PL, United Kingdom}
\affiliation{Department of Physics, University of Manchester, Manchester M13 9PL, UK}
\affiliation{Henry Royce Institute for Advanced Materials, Manchester M13 9PL, UK}
\author{Leonid I. Glazman}
\affiliation{Department of Physics, Yale University, New Haven, CT 06520, USA}

\date{\today}

\begin{abstract}
We develop the theory of collective modes supported by a Fermi liquid of electrons in pristine graphene. 
Under reasonable assumptions regarding the electron-electron interaction, all the modes but the plasmon are over-damped.
In addition to the $SU(2)$ symmetric spin mode, these include also the valley imbalance modes obeying a $U(1)$ symmetry, and a $U(2)$ symmetric valley spin imbalance mode.
We derive the interactions and diffusion constants characterizing the over-damped modes. The corresponding relaxation rates set fundamental constraints on graphene valley- and spintronics applications.
\end{abstract}
                             
\maketitle
\section{Introduction}
An extremely long electron mean free path~\cite{CastroNeto2009} combined with a fairly strong electron-electron interaction~\cite{Kotov2012} makes graphene an interesting platform for investigating Fermi liquid effects in solids.
The strength of interaction may be tuned by varying the electron density with respect to the neutrality point in graphene and by changing the dielectric environment encapsulating the graphene sheet.
A hallmark of Fermi liquid behavior is an emergence of the collective modes associated with the symmetries of the system.
The charge density mode, the plasmon, is the easiest to couple to, and can be probed in various spectroscopic experiments~\cite{Grimes1976}.
%and in DC transport measurements.
Plasmons are protected from Landau damping by their high propagation velocity, leading to fairly narrow spectroscopic lines.
The spectra of plasmons have been calculated for and measured in various settings~\cite{Grimes1976,Fetter1973,Mast1985}.

More recently, hydrodynamic electron flow has attracted much attention, occurring in the regime where the electron-electron mean free path is much smaller than other collision scales $\ell_{\text{e-e}} \ll \ell$. 
In the absence of disorder, a uniform electric current is protected by charge conservation and translation invariance, but a non-uniform electron flow is associated with size affects which arise from the electron viscosity.
To find it, one needs to solve the two-dimensional Fermi liquid kinetic equation~\cite{Lifshitz1980,Levitov2016} for the quasiparticle distribution function.
A variety of hydrodynamic size effects have been predicted and addressed experimentally in electric DC transport~\cite{Levitov2016,Bandurin2016,Sulpizio2019,Ku2020,Jenkins2020}.

A fairly weak spin-orbit interaction in carbon leaves electron spin in graphene approximately conserved.
Furthermore, the two Dirac points in graphene's electron spectrum are located far from each other in the Brillouin zone.
As the result, electron scattering between the valleys associated with the Dirac points is suppressed.
The approximate conservation of the spin and valley indices of a quasiparticle raises the question of the existence of sound modes in these channels.
It has also given rise to a host of proposals for ``spintronics'' and ``valleytronics'' graphene applications~\cite{Han2014,Schaibley2016} exploiting in various ways spin or valley currents.
The valleytronics proposals are specific for multi-valley materials, while the spintronics ones are a part of a broader semiconductor physics literature~\cite{Zutic2004,Avsar2020}.
In the analysis of spin- and valley-current conservation, the majority of theory works consider the effect of static disorder scattering non-interacting electrons at the Fermi level, see e.g. Refs.~\onlinecite{Mishchenko2004,Avsar2020}.
A notable exception is Ref.~\onlinecite{Flensberg2001} which evaluated in the Born approximation the spin relaxation rate due to electron-electron collisions in a two-dimensional electron gas, partially spin-polarized by a magnetic field.
Later this theory was applied to analyze the measurement of spin diffusion in the absence of polarizing magnetic field~\cite{Weber2005}.
We note in passing that the spin diffusion in the $SU(2)$-symmetric neutral three-dimensional Fermi liquid was considered in the context of the low-temperature He-3 properties~\cite{Leggett1968,Miyake1983}, and for ungapped graphene in 2D~\cite{Muller2011}.%but theory for a $SU(2)$-symmetric two-dimensional Fermi liquid apparently was not developed yet. 

In this work, we study dynamics of the neutral modes supported by the electron Fermi liquid in graphene.
The neutral modes include a $SU(2)$-symmetric spin mode, $U(1)$-symmetric modes of the inter-valley coherence and imbalance, as well as $U(2)$ symmetric inter-valley spin modes.
We identify and estimate the relative strength of the microscopic interactions which determine the values of parameters in a phenomenological Fermi liquid theory for these modes.
Under reasonable assumptions, all of the neutral modes are overdamped, and there is no neutral zero- or first-sound.
The spread and decay of the spin polarization density and of the inter-valley coherence and imbalance densities are thus characterized by their diffusion constants.
To find the diffusion constants, we evaluate the corresponding transport relaxation rates from the linearized collision integral  which accounts for the electron-electron scattering and solve the linearized kinetic equations.
The obtained rates depend on the temperature, electron density, and gaps at the Dirac points.
These gaps may appear due to deformation of the graphene lattice, for example in graphene encapsulated in hexagonal boron nitride~\cite{San-Jose2014}, and have a profound effect on electron backscattering due to the Berry flux redistribution across the Brillouin zone.
Indeed, backscattering of electrons in graphene may only occur in the presence of such gaps~\cite{Beenakker2008,Hwang2008}.

\section{Model}
A realistic level of the electron density induced by electrostatic gating corresponds to fairly small Fermi wave-vectors $k_F \ll \abs{\mathbf{K}}$ (measured from the respective Dirac points).
Therefore, we adopt the description of the electron system in terms of slowly-varying in space Fermi fields~\cite{Aleiner2007,Kharitonov2012}
\begin{equation}
  \hat{\Psi}_\sigma(\mathbf{r}) =
  \begin{psmallmatrix}
    u_{KA}(\mathbf{r})&
    u_{KB}(\mathbf{r})&
    u_{K'B}(\mathbf{r})&
    - u_{K'A}(\mathbf{r}) 
  \end{psmallmatrix}
  \cdot 
  \hat{\vec{\psi}}_{\sigma}(\mathbf{r})
  \label{eq:psi-slowly-varying}
\end{equation}
where $u_{\mathbf{k}\Sigma}(\mathbf{r})$ are the Bloch wavefunctions concentrated near the $\Sigma=A,B$ sub-lattice sites, at the $\mathbf{k}=\mathbf{K},\mathbf{K}'$ points in the Brillouin zone.
For definiteness, we consider the Fermi level above the Dirac point.
We may then perform a projection onto the conduction band $\vec{\psi}_{\mathbf{k}\sigma} = \sum_\zeta \chi_{\mathbf{k}\zeta} c_{\mathbf{k}\zeta\sigma}$ where the sub-lattice pseudo-spinors $\chi_{\mathbf{k},\zeta}$ are eigenvectors of the Dirac Hamiltonian,
\begin{equation}
 (v\mathbf{k} \cdot \bm{\Sigma} + \zeta \Delta \Sigma_3)\chi_{\mathbf{k}\zeta} = \sqrt{v^2k^2 + \Delta^2} \chi_{\mathbf{k}\zeta}
\end{equation}
with $v$ the Dirac velocity and $\Sigma_i$ are Pauli matrices in sub-lattice space~\footnote{Here we use $\Sigma$ for the sub-lattice Pauli matrices and reserve $\sigma$ for the spin Pauli matrices, as the sub-lattice degree of freedom will be projected out shortly. This differs from the notation commonly used in the literature of denoting the sub-lattice Pauli matrices $\sigma$ and spin Pauli matrices $s$.
We retain the use of $\tau$ for valley Pauli matrices.}
In terms of the upper-band operators, the low-energy Hamiltonian is
\begin{equation}
  \hat{H} =
  \sum_{\mathbf{k}\zeta\sigma} [\sqrt{v^2k^2 + \Delta^2}-(\Delta+E_F)] c^\dagger_{\mathbf{k}\zeta\sigma} c_{\mathbf{k}\zeta\sigma}
  + \hat{H}_\text{int}
\end{equation}
where $\zeta$ and $\sigma$ are respectively the valley and spin indices, $E_F$ is the Fermi level measured from the gap edge, and the interaction Hamiltonian is
\begin{multline}
  \hat{H}_\text{int} =
  \frac{1}{2} 
  \sum_{\mathbf{k}, \mathbf{k}', \mathbf{q}}
  \sum_{\substack{\zeta_1\zeta_1'\zeta_2\zeta_2'\\\sigma_1\sigma_1'\sigma_2\sigma_2'}}
    U^{\zeta_1\zeta_1';\zeta_2;\zeta'_2}_{\sigma_1\sigma_1';\sigma_2;\sigma_2}(\mathbf{k}, \mathbf{k}', \mathbf{q})\\
  \times
    :c^\dagger_{\zeta_1\sigma_1}(\mathbf{k} + \mathbf{q})
    c_{\zeta'_1\sigma'_1}(\mathbf{k})
    c^\dagger_{\zeta_2\sigma_2}(\mathbf{k}' - \mathbf{q})
    c_{\zeta'_2\sigma'_2}(\mathbf{k}'):
    \label{eq:Hint}
\end{multline}
where $:\cdots:$ denotes normal ordering with respect to the electron operators $c$.
The valley charge and spin symmetries constrain the short range interactions to be of the form (see~\cref{sec:sym-int})
\begin{multline}
U(\mathbf{p}, \mathbf{p}', \mathbf{q})
=
U^d _{\mathbf{p},\mathbf{p}',\mathbf{q}}
+ U^s_{\mathbf{p},\mathbf{p}',\mathbf{q}}
\bm{\sigma} \cdot \bm{\sigma}
\\
+ U^{v\parallel}_{\mathbf{p},\mathbf{p}',\mathbf{q}}
\bm{\tau}^\parallel \cdot \bm{\tau}^\parallel
+ U^{vz}_{\mathbf{p},\mathbf{p}',\mathbf{q}}
\tau^3 \tau^3
\\
 + U^{m\parallel}_{\mathbf{p},\mathbf{p}',\mathbf{q}}
\bm{\tau}^\parallel \cdot \bm{\tau}^\parallel
\bm{\sigma} \cdot \bm{\sigma}
 + U^{mz}_{\mathbf{p},\mathbf{p}',\mathbf{q}}
\tau^3 \tau^3
\bm{\sigma} \cdot \bm{\sigma},
\label{eq:gint}
\end{multline}
where $\tau^i$ and $\sigma^i$ are Pauli matrices in valley and spin space, respectively,
and all of the functions $U^{\alpha}$ are short ranged except for $U^d$ which includes the long range part of the Coulomb interaction, $V(q)$.
The six functions, $U^{\alpha}_{\mathbf{p}, \mathbf{p}', \mathbf{q}}$, are the inputs of our theory.
These, in turn, can be expressed in terms of the interaction constants $g_{zz},g_{\perp\perp},\widetilde g_{00}$ of an unprojected Hamiltonian~\cite{Kharitonov2012} and screened Coulomb potential $V(q)$, combined with the matrix elements of the projection operator constructed from eigenspinors $\chi_{\mathbf{k}\zeta}$. The latter contribute to the dependence of $U(\mathbf{p}, \mathbf{p}', \mathbf{q})$ on the respective momenta (cf.~\cref{sec:sym-int}).

The space-group symmetry of the graphene lattice constrains the form of the interaction Hamiltonian.
Allowing for the presence of a long-range density-density interaction, and neglecting the overlap of the Bloch functions on the $A$ and $B$ sub-lattices, the interaction Hamiltonian in terms of the unprojected operators $\psi_{k\zeta}$ in \cref{eq:psi-slowly-varying} takes the form~\footnotemark[\value{footnote}]
 (cf.~\cite{Kharitonov2012})
\begin{multline}
  \hat{H}^\psi_\text{int} = \frac{1}{2} \sum_{\mathbf{r},\mathbf{r}'}
  V(\mathbf{r} - \mathbf{r}') :\psi^\dagger(\mathbf{r}) \psi(\mathbf{r}) \psi^\dagger(\mathbf{r}') \psi(\mathbf{r}'): \\
  + \frac{1}{2} \sum_{\mathbf{r}}\sum_{\alpha\beta}
  \left[g_{\alpha\beta} :\psi^\dagger(\mathbf{r}) \Sigma^\alpha\tau^\beta \psi(\mathbf{r}) \psi^\dagger(\mathbf{r}') \Sigma^\alpha\tau^\beta \psi(\mathbf{r}'):\right.\\
  + \left.
   \tilde{g}_{00} :\psi^\dagger(\mathbf{r}) \Sigma^z\tau^z\psi(\mathbf{r}) \psi^\dagger(\mathbf{r}')\psi(\mathbf{r}'):\right],
   \label{eq:Hint-sublattice}
\end{multline}
with only $\tilde{g}_{00}$, $g_{zz}$, and $g_{\perp\perp} = g_{xx} = g_{xy} = g_{yx} = g_{yy}$ non-zero in the second term.
Projecting onto the upper bands reproduces the $U(2) \times U(2)$ symmetric form of the interaction \cref{eq:gint}.
The interaction functions in \cref{eq:gint} can then be expressed in terms of the interaction constants $g_{zz},g_{\perp\perp},\widetilde g_{00}$, and the function $V(q)$ combined with matrix elements of the eigenspinors $\chi_{\mathbf{k}\zeta}$ (see~\cref{sec:sym-int}).

One may estimate the interaction parameters here in terms of the matrix elements of the screened Coulomb interaction $V(q)$. To the lowest order one finds
\begin{equation}
\begin{gathered}
g_{\perp\perp}\sim V(\abs{\mathbf{K} - \mathbf{K}'}),\,\, g_{zz}\sim V(\abs{\mathbf{b}}),\\
\tilde{g}_{00}\sim \frac{g_{zz}\Delta}{v_F\abs{\mathbf{K} - \mathbf{K}'}}.
\end{gathered}
  \label{eq:interaction-harmonics}
\end{equation}
Like the Dirac gap $\Delta$, the constant $\tilde{g}_{00}$ is non-zero only if the lattice  $C_6$ symmetry is broken.
From the hierarchy of scales $q_{TF}, k_F \ll \abs{\mathbf{K} - \mathbf{K}'}$ we then have
\begin{equation}
 V(q \sim k_F) \gg g_{\perp\perp} > g_{zz} > \tilde{g}_{00}
 \label{eq:interaction-scale-hierarchy}
\end{equation}
(here $q_{TF}$ and $k_F$ are respectively the Thomas-Fermi and Fermi wavevectors).

\section{Landau-Fermi liquid theory of graphene away from the charge neutrality point}
The form of the interaction Hamiltonian \cref{eq:gint} stipulates an identical matrix form of the Landau functions $f(\mathbf{p} \cdot \mathbf{p}')$ of the phenomenological Fermi liquid theory. It also motivates the introduction of collective coordinates
\begin{equation}
\begin{gathered}
n(\mathbf{r}, \mathbf{p}) = \frac{1}{G_s G_v}\tr{\hat{\rho}(\mathbf{r}, \mathbf{p})}\\
\mathbf{s}(\mathbf{r}, \mathbf{p}) = \frac{1}{G_s G_v}\tr{\hat{\bm{\sigma}}\hat{\rho}(\mathbf{r}, \mathbf{p})}\\
\mathbf{Y}(\mathbf{r}, \mathbf{p}) = \frac{1}{G_s G_v}\tr{\hat{\bm{\tau}}\hat{\rho}(\mathbf{r}, \mathbf{p})}\\
M_{i}^j(\mathbf{r}, \mathbf{p}) = \frac{1}{G_s G_v}\tr{\hat{\tau}_i\hat{\sigma}_j\hat{\rho}(\mathbf{r}, \mathbf{p})}
\end{gathered}
    \label{eq:landau-densities}
\end{equation}
in terms of the Wigner transformed single particle density matrix $\hat{\rho}(\mathbf{r}, \mathbf{q})$, which may be interpreted as: charge $n$, spin $\mathbf{s}$, valley pseudo-spin $\mathbf{Y}$, and spin triplet valley pseudo-spin $\mathbf{M}_i$; $G_s$ and $G_v$ are spin and valley degeneracy, respectively.
In particular, as we will see below, the linear kinetic equation decouples in terms of these coordinates, such that each collective mode obeys an equation depending only on that channel.

For compactness, it is useful to define the arrays $\hat{X}^\mu,\rho^\mu,f^\mu$, with multi-index $\mu=(\alpha,\beta)$:
\begin{equation}
  \begin{gathered}
    \hat{X}^{\alpha\beta} = \hat{\tau}^\alpha\hat{\sigma}^\beta;\,\,\,\rho^{00} = n,\,\,\, \rho^{0i}= s_i,\,\,\, \rho^{i0}=Y_i,\,\,\,\rho^{ij} = M_i^j;\\
   f^{00} = f_d,\quad f^{0i}= f_s,\quad f^{i=1,2;0}=f_{v\parallel},\\
   f^{3;0} =  f_{vz},\quad f^{i=1,2;j}=f_{m\parallel},\quad f^{3;i} = f_{mz}.
  \end{gathered}
  \nonumber
\end{equation}
Here we have introduced the Landau functions $f_i(\mathbf{p} \cdot \mathbf{p}')$ for each channel. This allows us to write the Landau quasi-particle energy matrix as
\begin{multline}
  \hat{\epsilon}(\mathbf{p}, \mathbf{r}) = \xi(\mathbf{p})\hat{X}^{00}\\
  + \varphi^\mu(\mathbf{r})\hat{X}^\mu
  + \sum_{\mathbf{p}'}
    \sum_\mu \hat{X}^\mu f^\mu(\mathbf{p} \cdot \mathbf{p}')\hat \rho^\mu(\mathbf{r}, \mathbf{p'}),
    \label{eq:landau-qp-short}
\end{multline}
where $\xi(\mathbf{k})$ is the free particle excitation energy and $\varphi^\mu(\mathbf{r})$ is a generalized potential conjugate to $\rho^\mu$ (for the density channel this includes the self-consistent Vlasov field).

We may obtain linearized equations of motion for each of the collective coordinates from the Landau-Silin kinetic equation
\begin{equation}
  \frac{\partial \hat{\rho}}{\partial t}
  + \frac{1}{2}\left\{\frac{\partial \hat{\epsilon}}{\partial \mathbf{p}}, \frac{\partial \hat{\rho}}{\partial \mathbf{r}}\right\}
  - \frac{1}{2}\left\{\frac{\partial \hat{\epsilon}}{\partial \mathbf{r}}, \frac{\partial \hat{\rho}}{\partial \mathbf{p}}\right\}
  + i \left[\hat{\epsilon}, \hat{\rho}\right]
  = \hat{I}[\hat{\rho}]
  \label{eq:landau-silin}
\end{equation}
where $I[\hat{\rho}]$ is the collision integral.
We introduce the linearized deviation
$\hat{\rho}(\mathbf{k}, \mathbf{r}) = n_F(\bar{\epsilon}(\mathbf{k})) + \delta \hat{\rho}(\mathbf{k}, \mathbf{r})$,
where we have defined the self-consistently determined local equilibrium energy 
$\bar{\epsilon}(\mathbf{k}) = \epsilon[n_F(\bar{\epsilon}(\mathbf{k}))]$
via \cref{eq:landau-qp-short}. Expanding \cref{eq:landau-silin} to linear order in $\delta\rho$, one may take traces of the equation multiplied by the matrices $\hat{X}^\mu$, as in \cref{eq:landau-densities}, to obtain
\begin{multline}
  \frac{\partial \delta \rho^\mu(\mathbf{k}, \mathbf{r})}{\partial t}
  + \mathbf{v}\cdot \frac{\partial}{\partial \mathbf{r}}\delta\bar{\rho}^\mu(\mathbf{k}, \mathbf{r})
 + \left.\frac{\partial n}{\partial \epsilon}\right|_{\bar\epsilon}
 \mathbf{v}\cdot \bm{\mathcal{F}}^\mu
 \\
 = \frac{1}{G_s G_v}\tr\hat{X}^\mu\hat{I}[\delta \hat{\rho}].
    \label{eq:lin-lhs}
\end{multline}
with generalized forces $\bm{\mathcal{F}}^\mu = - \nabla \varphi^\mu$,
linearized deviations in each channel
\begin{equation}
\delta \rho^\mu(\mathbf{r}, \mathbf{p}) = \rho^\mu(\mathbf{r}, \mathbf{p}) - n_F(\bar{\epsilon}(\mathbf{k}))\delta_{\mu,00}
    \label{eq:linear-landau-densities}
\end{equation}
and deviation from local equlibrium~\cite{Nozieres1999}
\begin{multline}
 \delta\bar{\rho}^\mu (\mathbf{k}, \mathbf{r})= 
    \delta \rho^\mu(\mathbf{k}, \mathbf{r})\\
   - \left.\frac{\partial n}{\partial \epsilon}\right|_{\bar\epsilon}
   G_s G_v\sum_{\mathbf{p}}
   f^\mu(\mathbf{k} \cdot \mathbf{p})
   \delta \rho^\mu(\mathbf{p}, \mathbf{r}).
   \label{eq:rhobar}
\end{multline}
Note that while the Berry connection contributes to the full kinetic equation~\cite{Xiao2007a}, it does not enter the linearized equations of motion as long as we are interested in long-wavelength responses~\cite{Chen2017}.

At low temperatures $T \ll T_F$ the derivative of the Fermi function is sharply peaked at the Fermi level, pinning energies to the Fermi surface.
We therefore reparameterize the linearized deviations and Fermi liquid functions
\begin{equation}
 \delta \rho^\mu(\mathbf{p}, \mathbf{r}) \equiv -\left.\frac{\partial n}{\partial \epsilon}\right|_{\bar\epsilon}\nu^\mu(\phi,\mathbf{r})
\label{eq:nu-phi}
\end{equation}
in terms of the angular coordinate $\phi\, [\!\!\!\!\mod 2\pi]$ of $\mathbf{p}$ on the Fermi surface. 

We may now consider the collisionless limit of the kinetic equation \cref{eq:lin-lhs}, to determine whether undamped zero-sound modes exist. The charge-channel in this case gives rise to the usual zero-temperature 2D plasmon mode, which has been thoroughly studied~\cite{Stern1967,Fetter1973,Hwang2007,Lucas2018}, and thus we will focus here solely on the charge-neutral modes. Using the parametrization along with the real-space and time Fourier transforms in \cref{eq:lin-lhs}, we find for the collisionless ($\hat{I}[\delta \hat{\rho}]\to 0$) limit:
\begin{multline}
\omega \nu^\mu(\phi)
  - v_F q \cos\phi
  \Big[
    \nu^\mu(\phi)\\
 +
   \frac{G_s G_vp_F}{v_F}\oint {d\phi'}
   f^\mu(\phi - \phi')
   \nu^\mu(\phi')
    \Big]\\
    = v_F{\mathcal{F}}^\mu\cos(\phi-\chi).
    \label{eq:collisionless-ke}
\end{multline}
where $v_F$ is the Fermi velocity, $p_F$ the Fermi momentum, and $\phi$ and $\chi$ are the angles which the vectors  $\mathbf{p}$ and $\bm{\mathcal{F}}^\mu$ respectively make with $\mathbf{q}$.

We may estimate the Landau-Fermi liquid functions $f^\mu(\mathbf{p} \cdot \mathbf{p}')$ within the Hartree-Fock approximation. Due to the symmetrization of the interaction Hamiltonian one finds
\begin{equation}
 f^\mu(\mathbf{p} \cdot \mathbf{p}') \approx 2 U^\mu_{\mathbf{p}, \mathbf{p}', \mathbf{q}\to0}.
 \label{eq:hf-landau-function}
\end{equation}
Given the hierarchy of energy scales \cref{eq:interaction-scale-hierarchy}, the leading contribution to the Fermi-Liquid interaction functions comes from the long ranged-function $V(q)$,
\begin{multline}
  f^\mu(\theta) \approx
  -\frac{1}{2}V\cramped{\left[2k_F\sin\frac{\theta}{2}\right]}\\
  \times
  \left[\cos^2\left(\frac{\theta}{2}\right)
+ \frac{\Delta^2}{(\Delta+E_F)^2}\sin^2\left(\frac{\theta}{2}\right)\right]
\label{eq:fint-leading}
\end{multline}
where $\theta = \phi-\phi'$ is the angle between $\mathbf{p}$ and $\mathbf{p}'$.
We note that $f^\mu(\theta)<0$ at all $\theta$ for all neutral modes identified in \cref{eq:landau-densities}. Accounting for the smaller interaction constants identified in \cref{eq:interaction-harmonics} does not change this conclusion. At ${\mathcal{F}}^\mu=0$, we may bring \cref{eq:collisionless-ke} to the form
\begin{equation}
\widetilde\nu^\mu(\phi)=
   \frac{G_s G_v p_F}{v_F}\oint {d\phi'} f^\mu(\phi - \phi')\frac{\widetilde\nu^\mu(\phi')}{s-\cos\phi'}
   \label{eq:zero-sound}
\end{equation}
with $s=\omega/(v_Fq)$ and $\widetilde\nu^\mu=(s-\cos\phi)\nu^\mu$. Following the arguments of Ref.~\onlinecite{Lifsic2006}, we do not expect real-valued solutions with $s>1$ for $f^\mu<0$ (one may show that for constant interactions there are no solutions at all when $-1/2 < F^\mu_0 < 0$).
Therefore we infer that all neutral modes are overdamped~\cite{Klein2019}, albeit this does not exclude the possibility of a non-trivial response in the time domain~\cite{Klein2020}. 

\section{Relaxation of spin-valley collective modes}
The absence of propagating neutral modes leads us to conclude that at a finite temperature $T$ spreading of an initially-localized perturbation in these channels is ultimately controlled by diffusion.
In fact, as we will see, neutral first sound modes are not supported and the finite temperature behavior in these channels is generically diffusive.
Next, we evaluate the corresponding diffusion coefficients.
For that, we need to find transport scattering times $\tau^\mu_{tr}$. These are defined by the linearized collision integral in \cref{eq:lin-lhs}.
For temperatures below the Bloch-Gr\"uneisen temperature $T \ll T_{\text{BG}}$, electron-phonon scattering can be neglected as it will contribute at order $T^4$~\cite{Hwang2008,Hwang2009}.
Thus in a clean system, at low temperatures, the collision integral will be dominated by electron-electron collision which will be seen to contribute at order $T^2$.

The evaluation of the linearized collision integrals is greatly simplified by a choice of basis for each channel in which the linearized deviation of the density matrix is diagonal.
Because of the $SU(2)$ spin and $U(1)$ valley symmetries we need only consider the kinetic equations for $s^z,  Y_z, M_x^z$, and for $Y_x, M_z^z$. We therefore consider the collision integral in $\sigma_z\tau_z$ and $\sigma_z\tau_x$ bases, respectively. In each of the two, the linearized collision-integral matrix is diagonal and can be written in a familiar form
\begin{multline}
 I(\mathbf{p}_i, \alpha) =-\frac{1}{T} \sum_{\beta\gamma\delta} \sum_{\mathbf{p}_j \mathbf{p}_{i'} \mathbf{p}_{j'}}
 (2\pi)^2\delta(\cramped{\sum'_J} \mathbf{p}_{J})\\
 \times 2\pi\delta(\sum'_J \epsilon_J)n_i n_j(1-n_{i'})(1-n_{j'})\\
 \times
 W^{\alpha\beta;\gamma\delta}_{ij;i'j'}\left[\bar{\nu}_{i\alpha} + \bar{\nu}_{j\beta} - \bar{\nu}_{i'\gamma} - \bar{\nu}_{j'\delta}\right]
 \label{eq:schematic-collision}
 \end{multline}
in terms of the reparametrization
$\delta \bar\rho_{i\alpha}(\mathbf{p}, \mathbf{r}) \equiv -(\partial n/\partial \epsilon)\bar\nu_{i\alpha}(\phi,\mathbf{r})$
where $\bar{\rho}_{i\alpha} = \bar{\rho}_{\alpha\alpha}(\mathbf{p}_i)$ is the $\alpha=\sigma,\zeta$ component of the \emph{diagonal} deviation, defined analogously to \cref{eq:rhobar}, in the chosen basis. 
Here we have used the short-hand $\sum_J'h_J$ to denote sums of the form $h_i + h_j - h_{i'} - h_{j'}$
$n_i = n_F(\mathbf{p}_i)$ is the Fermi function at momentum $\mathbf{p}_i$, and $W$ is the square of the amplitude for two particles in states $\alpha,\beta$ to scatter into states $\gamma\delta$, which may be written in terms of the two particle $t$ matrix as $ W^{\alpha\beta;\gamma\delta}_{ij;ij'}= \left|\braket{i\alpha;j\beta|\hat{t}|i'\gamma;j'\delta}\right|^2$.

The trace operation on the right-hand of \cref{eq:lin-lhs} is then simply the weighted sum $\tr(\hat X^\mu\hat I) = \sum_\alpha \lambda_\alpha^\mu I_\alpha$, where $\lambda_\alpha^\mu$ is the eigenvalue corresponding to eigenvector $|\alpha\rangle$ of matrix $X^\mu$.
Upon performing the trace the collision integral for each channel can be separated into two parts: those involving scattering of particles with the same quantum number in that channel and those involving scattering of particles with different quantum numbers (e.g.\ same spin or opposite spin respectively for the case of the spin mode, cf. \cref{eq:scattering-rates}), which we denote $I_+$ and $I_-$, respectively.
The former has the same structure as the collision integral for the charge channel and is known in 2D to give rise to a transport scattering time which goes as $T^4$ for low temperatures~\cite{Ledwith2019}.
The latter term on the other hand will be seen to scale as $T^2\ln T$ and comprises the dominant contribution to relaxation of currents in each channel,
\begin{multline}
  \tr(\hat X^\mu\hat I)\to I^\mu_-(\mathbf{p}_i)
  = -\frac{1}{T} 
 \sum_{\mathbf{p}_j \mathbf{p}_{i'} \mathbf{p}_{j'}}
 (2\pi)^2\delta\left(\cramped{\sum'_J \mathbf{p}_{J}}\right)\\
 \times 2\pi\delta\left(\cramped{\sum'_J \epsilon_J}\right)n_i n_j(1-n_{i'})(1-n_{j'})\\ \times
 W^\mu_- \left(
\bar{\nu}^\mu_{i} -  \bar{\nu}^\mu_{j} -  \bar{\nu}^\mu_{i'} + \bar{\nu}^\mu_{j'}\right)
 \label{eq:Imuminus}
\end{multline}
with scattering probabilities
\begin{equation}
  \begin{aligned}
 W^s_- &= 2(W^D_{\uparrow\downarrow;++}  + W^D_{\uparrow\downarrow;+-})\\
 W^{vz}_- &= 2(W^D_{\uparrow\uparrow;+-} + W^D_{\uparrow\downarrow;+-})\\
 W^{mz}_- &= 2(W^D_{\uparrow\uparrow;+-} + W^D_{\uparrow\downarrow;++})\\
 W^{v\parallel}_- &= 2(W^{xD}_{\uparrow\uparrow;+-} + W^{xD}_{\uparrow\downarrow;+-})\\
 W^{m\parallel}_- &= 2(W^{xD}_{\uparrow\uparrow;+-} + W^{xD}_{\uparrow\downarrow;++})
\end{aligned}
\label{eq:scattering-rates}
\end{equation}
where $W^D_{\sigma\sigma';\zeta\zeta'}$ is the scattering probability for two \emph{distinguishable} particles with spins $\sigma,\sigma'$ and valley indices $\zeta,\zeta'$ and the superscript $x$ indicates the choice of the $\tau^x$ eigen-basis in valley space.
These may in turn be written in terms of components of the $t$-matrix.
In the first Born approximation, this can be expressed in terms of the functions in \cref{eq:gint} as
\begin{equation}
  \begin{aligned}
  W^D_{\uparrow\downarrow;++} &= 4\abs{U_d - U_s + U_{vz} - U_{mz}}^2\\
  W^D_{\uparrow\downarrow;+-}&= 4\abs{U_d - U_s - U_{vz} + U_{mz}}^2\\
  W^D_{\uparrow\uparrow;+-}&= 4\abs{U_d + U_s - U_{vz} - U_{mz}}^2\\
  W^{xD}_{\uparrow\downarrow;++}&= 4\abs{U_d - U_s + U_{v\parallel} - U_{m\parallel}}^2\\
  W^{xD}_{\uparrow\downarrow;+-}&= 4\abs{U_d - U_s - U_{v\parallel} + U_{m\parallel}}^2\\
  W^{xD}_{\uparrow\uparrow;+-}&= 4\abs{U_d + U_s - U_{v\parallel} - U_{m\parallel}}^2
\end{aligned}
\label{eq:Born}
\end{equation}

To the lowest order in $T/T_F$, the Fermi functions restrict the summation over momenta in \cref{eq:Imuminus} to the states close to the Fermi surface. One may then transform~\cite{Nozieres1999,Ledwith2017} the summation to integration over the energies $\epsilon,\epsilon^\prime$ of the incoming particles, the energy transferred in a collision $\omega$, and the scattering angle $\theta_\text{sc}$. Due to the constraints on $\epsilon,\epsilon^\prime$ and conservation laws, the incoming particles collide almost head-on, or their momenta are almost collinear to each other.
To evaluate the transport relaxation times $\tau^\mu_\text{tr}$, we use $\nu^\mu(\phi)\propto\cos\phi$~\footnote{Because of the time-reversal symmetry of the system this is equivalent to calculating $\tau^\mu_\text{tr} \equiv \tau^\mu_1$.} to arrive at
\begin{multline}
 \frac{1}{\tau^\mu_{\text{tr}}}  =
  \frac{4\nu_F T^2}{v_F^2}
  \int d\Sigma
  \int_0^{\pi-\theta_c} \frac{d\theta_\text{sc}}{\sin\theta_\text{sc}}(1-\cos  \theta_\text{sc})\\
  \times \left(
   W^\mu_{-, \text{collinear}}(\theta_\text{sc}) +  W^\mu_{-, \text{head-on}}(\theta_\text{sc})\right).
  \label{eq:relax-tr-sc}
\end{multline}
The logarithmic divergence~\cite{Menashe1996} at $\theta_{\text{sc}} = \pi$ is cut off by $\theta_c\sim T/E_F$ due to the kinematic constraints on scattering of particles on the Fermi surface (see \cref{sec:backscatter} for details).
For compactness, we have defined the dimensionless energy integration measure
\begin{equation}
 d\Sigma =
  \frac{1}{4\pi^2} 
  du du' dw
n_i n_j (1 - n_{i'}) (1-n_{j'})
  \label{eq:energy-measure}
\end{equation}
in terms of dimensionless variables $u=\epsilon/T$, $w=\omega/T$, $\epsilon_{i},\epsilon_{i'} = \epsilon \pm \omega/2$, $\epsilon_{j},\epsilon_{j'} = \epsilon' \mp \omega/2$.
In the case where $q_{TF}\ll 2k_F$ there is also a logarithmic contribution due to collinear scattering by the Coulomb potential, which is cut off by the Thomas-Fermi wavenumber $q_{TF}$~\footnotetext{We perform calculations in the static screening approximation. Dynamic screening effects are unimportant so long as we remain at low temperatures $v q_{TF} \gg T$ (see \cref{sec:forward}), but see Ref.~\onlinecite{Zala2001,*Narozhny2002} and the supplement to Ref.~\onlinecite{Alekseev2020} for discussion of the effects of dynamic screening.}\cite{Note3,Zala2001,Narozhny2002,Alekseev2020}.
In graphene, unlike more conventional Fermi liquids, $q_{TF} = G_s G_V \alpha v k_F /v_F$ can be greater than $2k_F$ due to the degeneracy factors and strong effective coupling $\alpha=e^2/\kappa v$.
In this complementary regime, the collinear scattering logarithm is absent and the dominant matrix element for backscattering will be screened.
We present here explicit expressions for the former case, $q_{TF} \ll 2 k_F$, but one may straightforwardly perform the analogous calculations in the latter case and the qualitative results remain the same.
Which regime is realized experimentally will depend on the background dielectric constant and doping.

Thus, performing the integration in \cref{eq:relax-tr-sc}, with $q_{TF} \ll 2 k_F$, we find
\begin{equation}
\begin{aligned}
\frac{1}{\tau^\mu_{\rm tr}(T)}&=\frac{4\pi}{3}(E_F+\Delta)\alpha^2\frac{T^2}{E_F^2}\left(\frac{1+E_F/\Delta}{1+E_F/2\Delta}\right)^{\!2}\\
&\times\left\{\left(\frac{\Delta}{\Delta + E_F}\right)^{\!4}\ln\frac{\sqrt{E_F(E_F\!+2\Delta)}}{T}\right.\\
&\left.+\ln\frac{\sqrt{E_F(E_F\!+2\Delta)}}{v q_{TF}}\right\}
\label{eq:taudelta}
\end{aligned}
\end{equation}
 in the leading logarithmic approximation, applicable at $T\ll\sqrt{E_F\Delta}$.
 In the evaluation of the scattering probabilities entering \cref{eq:relax-tr-sc}, we used the Born approximation and the Thomas-Fermi screened Coulomb potential, cf. \cref{eq:interaction-scale-hierarchy,eq:Born}. The first logarithmic contribution in \cref{eq:taudelta} comes from the backscattering amplitude ($\theta_{\rm sc}=\pi$), as long as it remains finite. Notably, backscattering is suppressed in graphene at $\Delta\ll E_F$ due to the presence of Berry phase in the electron eigenfunctions~\cite{Beenakker2008,Hwang2008}.
 Therefore, in the relativistic limit $\Delta\to 0$, \cref{eq:taudelta} is replaced by (see \cref{sec:Wback})
 \begin{equation}
 \begin{aligned}
 \frac{1}{\tau^\mu_{\rm tr}(T)}&=\frac{4}{3\pi}\frac{T^2}{E_F^2}\left\{N_\mu\left(g_\mu\frac{E_F}{v_F}\right)^{\!2}\ln\frac{E_F}{T}\right.\\
&+\left.\left(2\pi\alpha\right)^2E_F\ln\frac{E_F}{vq_{TF}}\right\}.
 \end{aligned}
 \label{eq:taunodelta}
 \end{equation}
 Here $g_\mu=g_{\perp\perp}$ for $\mu=vz,mz,v\!\parallel,m\!\parallel$, and $g_\mu=g_{zz}$ for $\mu=s$, and the numerical factor $N_\mu=8$ for $\mu=vz,mz$, $N_\mu=10$ for $\mu=v\!\!\parallel, m\!\!\parallel$, and $N_\mu=4$ for $\mu=s$. We note that the logarithmic terms associated with the backscattering in \cref{eq:taunodelta} generically are smaller than those present at $\Delta\sim E_F$, cf. \cref{eq:interaction-harmonics,eq:interaction-scale-hierarchy,eq:taudelta}. The diffusion constant for each of the channels is $D_\mu=v_F^2 \tau^\mu_{\text{tr}}/2$ (consistent with the Born approximation, here we dispensed with the Fermi liquid correction~\cite{Lifsic2006,Nozieres1999} to the Fermi velocity). The diffusion regime settles in at times $t\gtrsim \tau_{\rm tr}(T)$.

\section{Discussion}
The relaxation rates for the neutral modes can be experimentally probed through non-local resistance measurements using the  spin- and valley- Hall effects~\cite{Xiao2007a,Abanin2009,Yamamoto2015}.
Currently experimental measurements of spin and valley diffusion below the Bloch-Gr\"ueisen temperature have obtained diffusion constants corresponding to mean free paths of the order $\SI{0.1}{\micro\meter}$,~\cite{Gorbachev2014,Ingla-Aynes2015,Yamamoto2015,Gurram2018}.
Calculations performed in the limit $q_{TF} \ll k_F$ give significantly longer mean free paths, indicating either that impurities play the dominant role, or the system is in the regime $k_F \ll q_{TF}$.
Nonetheless, experimental works on electron hydrodynamics suggest that it could be possible to reach the regime where electron-electron effects dominate the response of the neutral modes~\cite{Levitov2016,Bandurin2016,Sulpizio2019,Ku2020,Jenkins2020}, as considered here. 
Thus, the predicted relaxation rates -- and their temperature dependence -- should be measurable, either via the same types of experiments as have been previously used to measure spin and valley diffusion, or through other methods~\cite{Liu2020}.

The generalization of \cref{eq:relax-tr-sc} to higher angular harmonics ($m\geq 2$) of the distribution function $\nu(\phi)$ is presented in~\cref{sec:relax-arb}.
The relaxation rate $1/\tau_2(T)$ of the $m=2$ harmonic is similar to $1/\tau_{\rm tr}(T)$ of \cref{eq:taudelta,eq:taunodelta} and is likely lower than $1/\tau_{\rm tr}(T)$ in the respective limits, as $1/\tau_2(T)$ lacks the logarithmic enhancement of the backscattering contribution.

The hierarchy $\tau_{\rm tr}\lesssim\tau_2$ for all of the graphene Fermi liquid neutral modes excludes the possibility of a hydrodynamic sound mode, contrary to the case of the density mode in a conventional neutral Fermi liquid (for the density mode, $1/\tau_{\rm tr}\equiv 0$ by translation invariance). 
Combined with the discussion below \cref{eq:zero-sound}, we thus find that both zero and first sound are absent in \emph{all neutral channels}.

% In general the even $m$ relaxation rates asymptotically grow as $\log m$ up to $m_\text{max} \sim \sqrt{T_F/T}$~\cite{Lee2016,Ledwith2017,Ledwith2019}, for both the charged and neutral modes.
% As a consequence, the transverse spreading of tightly focused beams, such as in transverse electron focusing (TEF), is determined by the behavior of the large $m$ rates, which may be written in the form $\tau_{m\gg2k_F/q_{TF}}^{-1}\sim (T^2/T_*) \log m$.
% This allows for the characteristic energy $T_*$ of the pre-logarithmic factor to be extracted from the temperature and contact separation dependent decay of TEF signals~\cite{Lee2016}.
% Since, the large $m$ modes in all channels share an energy scale $T_*\propto V(0)^{-2} = (G_s G_v [1 + F_0^d])^2$ this provides a means to experimentally determine the relaxation rates for the neutral spin-valley collective modes.
Relaxation of the higher-$m$ moments of the distribution function can be measured in magnetic focusing experiments~\cite{Lee2016}. With the increase of $m$, the role of the forward-scattering contribution (which gave rise to the second term in \cref{eq:taudelta,eq:taunodelta}) strengthens. At $m\gg 2k_F/q_{TF}$ and sufficiently low temperatures the small-angle scattering involving the screened Coulomb potential  $V(0)= 1/(G_s G_v [1 + F_0^d])$ dominates the relaxation rate $1/\tau_m\sim (T^2/v k_F) \ln m$. The asymptotic large-$m$ relaxation of spin modes can then be accessed in focusing experiments utilizing spin-polarized leads in a setup similar to that of Ref.~\onlinecite{Lee2016}.

In the above we have assumed $SU(2)$ spin invariance, leading to diffusion of the conserved spin density as described by \cref{eq:taudelta,eq:taunodelta}.
Spin-orbit coupling destroys the $SU(2)$ symmetry and leads to the interaction-induced relaxation of a net spin polarization.
An expression for the associated relaxation rate was obtained by \citet{Glazov2002} in terms of the spin-orbit coupling strength and electron-electron collision rates~\cite{*[{See Eq. (25) in }] [{}] Glazov2004}. 
We neglect such a combined effect here as intrinsic spin-orbit coupling in graphene is weak~\cite{Han2014}, but in the presence of extrinsic spin-orbit coupling~\cite{*[{See for example, }] [{}] Wang2016c} similar relaxation effects would be expected for the spin-valley channels studied in this work.
 
 Throughout this work we disregarded the trigonal warping of the electron spectrum in graphene. Warping does not destroy the used $U(2) \times U(2)$ spin-valley symmetries.
 The modification of the spectrum, however, cuts off the logarithmic singularity of backscattering in \cref{eq:relax-tr-sc} and introduces anisotropy of the diffusion coefficient due to the dependence of the Fermi velocity on the direction of the electron wavevector.
 For similar reasons, the calculations presented herein may be extended to twisted bilayer graphene, which also possesses an internal $U(2) \times U(2)$ symmetry~\cite{Tarnopolsky2019,Vafek2020}, and thus at a generic filling we also expect similar qualitative behavior such as the absence of neutral zero or first sound, and a transport scattering time for the neutral modes which scales as $(W/T)^2 \ln W/T$ where $W$ is the bandwidth of the nearly-flat bands.
 As the flat-band limit is approached one must consider both the valence and conduction bands together and modes associated with the inter-band transitions will appear which could exhibit different behaviors given the enlarged symmetry $U(4)$ ($U(4) \times U(4)$ for the chiral case)~\cite{Tarnopolsky2019,Vafek2020} of the flat-band limit.

\begin{acknowledgments}
The authors would like to thank M. Kharitonov and I. Aleiner. ZR would also like to thank J. Wilson for helpful discussions.
This work was supported by NSF DMR-2002275 (LG), European Graphene Flagship Core 3 Project, Lloyd Register Foundation Nanotechnology Grant, EPSRC grants EP/S030719/1 and EP/N010345/1 (VF), and the Yale Prize Postdoctoral Fellowship in Condensed Matter Theory (ZR).
\end{acknowledgments}

%% CAREFUL: BIBTEX FILE IS AUTOGENERATED
\bibliography{references.gen}

\begin{thebibliography}{61}%
\makeatletter
\providecommand \@ifxundefined [1]{%
 \@ifx{#1\undefined}
}%
\providecommand \@ifnum [1]{%
 \ifnum #1\expandafter \@firstoftwo
 \else \expandafter \@secondoftwo
 \fi
}%
\providecommand \@ifx [1]{%
 \ifx #1\expandafter \@firstoftwo
 \else \expandafter \@secondoftwo
 \fi
}%
\providecommand \natexlab [1]{#1}%
\providecommand \enquote  [1]{``#1''}%
\providecommand \bibnamefont  [1]{#1}%
\providecommand \bibfnamefont [1]{#1}%
\providecommand \citenamefont [1]{#1}%
\providecommand \href@noop [0]{\@secondoftwo}%
\providecommand \href [0]{\begingroup \@sanitize@url \@href}%
\providecommand \@href[1]{\@@startlink{#1}\@@href}%
\providecommand \@@href[1]{\endgroup#1\@@endlink}%
\providecommand \@sanitize@url [0]{\catcode `\\12\catcode `\$12\catcode
  `\&12\catcode `\#12\catcode `\^12\catcode `\_12\catcode `\%12\relax}%
\providecommand \@@startlink[1]{}%
\providecommand \@@endlink[0]{}%
\providecommand \url  [0]{\begingroup\@sanitize@url \@url }%
\providecommand \@url [1]{\endgroup\@href {#1}{\urlprefix }}%
\providecommand \urlprefix  [0]{URL }%
\providecommand \Eprint [0]{\href }%
\providecommand \doibase [0]{https://doi.org/}%
\providecommand \selectlanguage [0]{\@gobble}%
\providecommand \bibinfo  [0]{\@secondoftwo}%
\providecommand \bibfield  [0]{\@secondoftwo}%
\providecommand \translation [1]{[#1]}%
\providecommand \BibitemOpen [0]{}%
\providecommand \bibitemStop [0]{}%
\providecommand \bibitemNoStop [0]{.\EOS\space}%
\providecommand \EOS [0]{\spacefactor3000\relax}%
\providecommand \BibitemShut  [1]{\csname bibitem#1\endcsname}%
\let\auto@bib@innerbib\@empty
%</preamble>
\bibitem [{\citenamefont {Castro~Neto}\ \emph {et~al.}(2009)\citenamefont
  {Castro~Neto}, \citenamefont {Guinea}, \citenamefont {Peres}, \citenamefont
  {Novoselov},\ and\ \citenamefont {Geim}}]{CastroNeto2009}%
  \BibitemOpen
  \bibfield  {author} {\bibinfo {author} {\bibfnamefont {A.~H.}\ \bibnamefont
  {Castro~Neto}}, \bibinfo {author} {\bibfnamefont {F.}~\bibnamefont {Guinea}},
  \bibinfo {author} {\bibfnamefont {N.~M.~R.}\ \bibnamefont {Peres}}, \bibinfo
  {author} {\bibfnamefont {K.~S.}\ \bibnamefont {Novoselov}},\ and\ \bibinfo
  {author} {\bibfnamefont {A.~K.}\ \bibnamefont {Geim}},\ }\bibfield  {title}
  {\bibinfo {title} {The electronic properties of graphene},\ }\href
  {https://doi.org/10.1103/revmodphys.81.109} {\bibfield  {journal} {\bibinfo
  {journal} {Rev. Mod. Phys.}\ }\textbf {\bibinfo {volume} {81}},\ \bibinfo
  {pages} {109} (\bibinfo {year} {2009})}\BibitemShut {NoStop}%
\bibitem [{\citenamefont {Kotov}\ \emph {et~al.}(2012)\citenamefont {Kotov},
  \citenamefont {Uchoa}, \citenamefont {Pereira}, \citenamefont {Guinea},\ and\
  \citenamefont {Castro~Neto}}]{Kotov2012}%
  \BibitemOpen
  \bibfield  {author} {\bibinfo {author} {\bibfnamefont {V.~N.}\ \bibnamefont
  {Kotov}}, \bibinfo {author} {\bibfnamefont {B.}~\bibnamefont {Uchoa}},
  \bibinfo {author} {\bibfnamefont {V.~M.}\ \bibnamefont {Pereira}}, \bibinfo
  {author} {\bibfnamefont {F.}~\bibnamefont {Guinea}},\ and\ \bibinfo {author}
  {\bibfnamefont {A.~H.}\ \bibnamefont {Castro~Neto}},\ }\bibfield  {title}
  {\bibinfo {title} {Electron-{{Electron Interactions}} in {{Graphene}}:
  {{Current Status}} and {{Perspectives}}},\ }\href
  {https://doi.org/10.1103/revmodphys.84.1067} {\bibfield  {journal} {\bibinfo
  {journal} {Rev. Mod. Phys.}\ }\textbf {\bibinfo {volume} {84}},\ \bibinfo
  {pages} {1067} (\bibinfo {year} {2012})}\BibitemShut {NoStop}%
\bibitem [{\citenamefont {Grimes}\ and\ \citenamefont
  {Adams}(1976)}]{Grimes1976}%
  \BibitemOpen
  \bibfield  {author} {\bibinfo {author} {\bibfnamefont {C.~C.}\ \bibnamefont
  {Grimes}}\ and\ \bibinfo {author} {\bibfnamefont {G.}~\bibnamefont {Adams}},\
  }\bibfield  {title} {\bibinfo {title} {Observation of {{Two}}-{{Dimensional
  Plasmons}} and {{Electron}}-{{Ripplon Scattering}} in a {{Sheet}} of
  {{Electrons}} on {{Liquid Helium}}},\ }\href
  {https://doi.org/10.1103/physrevlett.36.145} {\bibfield  {journal} {\bibinfo
  {journal} {Phys. Rev. Lett.}\ }\textbf {\bibinfo {volume} {36}},\ \bibinfo
  {pages} {145} (\bibinfo {year} {1976})}\BibitemShut {NoStop}%
\bibitem [{\citenamefont {Fetter}(1973)}]{Fetter1973}%
  \BibitemOpen
  \bibfield  {author} {\bibinfo {author} {\bibfnamefont {A.~L.}\ \bibnamefont
  {Fetter}},\ }\bibfield  {title} {\bibinfo {title} {Electrodynamics of a
  layered electron gas. {{I}}. {{Single}} layer},\ }\href
  {https://doi.org/10.1016/0003-4916(73)90161-9} {\bibfield  {journal}
  {\bibinfo  {journal} {Ann. Phys.}\ }\textbf {\bibinfo {volume} {81}},\
  \bibinfo {pages} {367} (\bibinfo {year} {1973})}\BibitemShut {NoStop}%
\bibitem [{\citenamefont {Mast}\ \emph {et~al.}(1985)\citenamefont {Mast},
  \citenamefont {Dahm},\ and\ \citenamefont {Fetter}}]{Mast1985}%
  \BibitemOpen
  \bibfield  {author} {\bibinfo {author} {\bibfnamefont {D.~B.}\ \bibnamefont
  {Mast}}, \bibinfo {author} {\bibfnamefont {A.~J.}\ \bibnamefont {Dahm}},\
  and\ \bibinfo {author} {\bibfnamefont {A.~L.}\ \bibnamefont {Fetter}},\
  }\bibfield  {title} {\bibinfo {title} {Observation of {{Bulk}} and {{Edge
  Magnetoplasmons}} in a {{Two}}-{{Dimensional Electron Fluid}}},\ }\href
  {https://doi.org/10.1103/physrevlett.54.1706} {\bibfield  {journal} {\bibinfo
   {journal} {Phys. Rev. Lett.}\ }\textbf {\bibinfo {volume} {54}},\ \bibinfo
  {pages} {1706} (\bibinfo {year} {1985})}\BibitemShut {NoStop}%
\bibitem [{\citenamefont {Lifshitz}\ and\ \citenamefont
  {Pitaevski}(1980)}]{Lifshitz1980}%
  \BibitemOpen
  \bibfield  {author} {\bibinfo {author} {\bibfnamefont {E.~M.}\ \bibnamefont
  {Lifshitz}}\ and\ \bibinfo {author} {\bibfnamefont {L.~P.}\ \bibnamefont
  {Pitaevski}},\ }\bibfield  {title} {\bibinfo {title} {The {{Normal Fermi
  Liquid}}},\ }in\ \href {https://doi.org/10.1016/B978-0-08-050350-9.50006-X}
  {\emph {\bibinfo {booktitle} {Statistical {{Physics}}}}},\ \bibinfo {editor}
  {edited by\ \bibinfo {editor} {\bibfnamefont {E.~M.}\ \bibnamefont
  {Lifshitz}}\ and\ \bibinfo {editor} {\bibfnamefont {L.~P.}\ \bibnamefont
  {Pitaevski}}}\ (\bibinfo  {publisher} {{Butterworth-Heinemann}},\ \bibinfo
  {address} {{Oxford}},\ \bibinfo {year} {1980})\ pp.\ \bibinfo {pages}
  {1--27}\BibitemShut {NoStop}%
\bibitem [{\citenamefont {Levitov}\ and\ \citenamefont
  {Falkovich}(2016)}]{Levitov2016}%
  \BibitemOpen
  \bibfield  {author} {\bibinfo {author} {\bibfnamefont {L.}~\bibnamefont
  {Levitov}}\ and\ \bibinfo {author} {\bibfnamefont {G.}~\bibnamefont
  {Falkovich}},\ }\bibfield  {title} {\bibinfo {title} {Electron viscosity,
  current vortices and negative nonlocal resistance in graphene},\ }\href
  {https://doi.org/10.1038/nphys3667} {\bibfield  {journal} {\bibinfo
  {journal} {Nat. Phys.}\ }\textbf {\bibinfo {volume} {12}},\ \bibinfo {pages}
  {672} (\bibinfo {year} {2016})}\BibitemShut {NoStop}%
\bibitem [{\citenamefont {Bandurin}\ \emph {et~al.}(2016)\citenamefont
  {Bandurin}, \citenamefont {Torre}, \citenamefont {Krishna~Kumar},
  \citenamefont {Ben~Shalom}, \citenamefont {Tomadin}, \citenamefont
  {Principi}, \citenamefont {Auton}, \citenamefont {Khestanova}, \citenamefont
  {Novoselov}, \citenamefont {Grigorieva}, \citenamefont {Ponomarenko},
  \citenamefont {Geim},\ and\ \citenamefont {Polini}}]{Bandurin2016}%
  \BibitemOpen
  \bibfield  {author} {\bibinfo {author} {\bibfnamefont {D.~A.}\ \bibnamefont
  {Bandurin}}, \bibinfo {author} {\bibfnamefont {I.}~\bibnamefont {Torre}},
  \bibinfo {author} {\bibfnamefont {R.}~\bibnamefont {Krishna~Kumar}}, \bibinfo
  {author} {\bibfnamefont {M.}~\bibnamefont {Ben~Shalom}}, \bibinfo {author}
  {\bibfnamefont {A.}~\bibnamefont {Tomadin}}, \bibinfo {author} {\bibfnamefont
  {A.}~\bibnamefont {Principi}}, \bibinfo {author} {\bibfnamefont {G.~H.}\
  \bibnamefont {Auton}}, \bibinfo {author} {\bibfnamefont {E.}~\bibnamefont
  {Khestanova}}, \bibinfo {author} {\bibfnamefont {K.~S.}\ \bibnamefont
  {Novoselov}}, \bibinfo {author} {\bibfnamefont {I.~V.}\ \bibnamefont
  {Grigorieva}}, \bibinfo {author} {\bibfnamefont {L.~A.}\ \bibnamefont
  {Ponomarenko}}, \bibinfo {author} {\bibfnamefont {A.~K.}\ \bibnamefont
  {Geim}},\ and\ \bibinfo {author} {\bibfnamefont {M.}~\bibnamefont {Polini}},\
  }\bibfield  {title} {\bibinfo {title} {Negative local resistance caused by
  viscous electron backflow in graphene.},\ }\href
  {https://doi.org/10.1126/science.aad0201} {\bibfield  {journal} {\bibinfo
  {journal} {Science}\ }\textbf {\bibinfo {volume} {351}},\ \bibinfo {pages}
  {1055} (\bibinfo {year} {2016})}\BibitemShut {NoStop}%
\bibitem [{\citenamefont {Sulpizio}\ \emph {et~al.}(2019)\citenamefont
  {Sulpizio}, \citenamefont {Ella}, \citenamefont {Rozen}, \citenamefont
  {Birkbeck}, \citenamefont {Perello}, \citenamefont {Dutta}, \citenamefont
  {{Ben-Shalom}}, \citenamefont {Taniguchi}, \citenamefont {Watanabe},
  \citenamefont {Holder}, \citenamefont {Queiroz}, \citenamefont {Principi},
  \citenamefont {Stern}, \citenamefont {Scaffidi}, \citenamefont {Geim},\ and\
  \citenamefont {Ilani}}]{Sulpizio2019}%
  \BibitemOpen
  \bibfield  {author} {\bibinfo {author} {\bibfnamefont {J.~A.}\ \bibnamefont
  {Sulpizio}}, \bibinfo {author} {\bibfnamefont {L.}~\bibnamefont {Ella}},
  \bibinfo {author} {\bibfnamefont {A.}~\bibnamefont {Rozen}}, \bibinfo
  {author} {\bibfnamefont {J.}~\bibnamefont {Birkbeck}}, \bibinfo {author}
  {\bibfnamefont {D.~J.}\ \bibnamefont {Perello}}, \bibinfo {author}
  {\bibfnamefont {D.}~\bibnamefont {Dutta}}, \bibinfo {author} {\bibfnamefont
  {M.}~\bibnamefont {{Ben-Shalom}}}, \bibinfo {author} {\bibfnamefont
  {T.}~\bibnamefont {Taniguchi}}, \bibinfo {author} {\bibfnamefont
  {K.}~\bibnamefont {Watanabe}}, \bibinfo {author} {\bibfnamefont
  {T.}~\bibnamefont {Holder}}, \bibinfo {author} {\bibfnamefont
  {R.}~\bibnamefont {Queiroz}}, \bibinfo {author} {\bibfnamefont
  {A.}~\bibnamefont {Principi}}, \bibinfo {author} {\bibfnamefont
  {A.}~\bibnamefont {Stern}}, \bibinfo {author} {\bibfnamefont
  {T.}~\bibnamefont {Scaffidi}}, \bibinfo {author} {\bibfnamefont {A.~K.}\
  \bibnamefont {Geim}},\ and\ \bibinfo {author} {\bibfnamefont
  {S.}~\bibnamefont {Ilani}},\ }\bibfield  {title} {\bibinfo {title}
  {Visualizing {{Poiseuille}} flow of hydrodynamic electrons.},\ }\href
  {https://doi.org/10.1038/s41586-019-1788-9} {\bibfield  {journal} {\bibinfo
  {journal} {Nature}\ }\textbf {\bibinfo {volume} {576}},\ \bibinfo {pages}
  {75} (\bibinfo {year} {2019})}\BibitemShut {NoStop}%
\bibitem [{\citenamefont {Ku}\ \emph {et~al.}(2020)\citenamefont {Ku},
  \citenamefont {Zhou}, \citenamefont {Li}, \citenamefont {Shin}, \citenamefont
  {Shi}, \citenamefont {Burch}, \citenamefont {Anderson}, \citenamefont
  {Pierce}, \citenamefont {Xie}, \citenamefont {Hamo}, \citenamefont {Vool},
  \citenamefont {Zhang}, \citenamefont {Casola}, \citenamefont {Taniguchi},
  \citenamefont {Watanabe}, \citenamefont {Fogler}, \citenamefont {Kim},
  \citenamefont {Yacoby},\ and\ \citenamefont {Walsworth}}]{Ku2020}%
  \BibitemOpen
  \bibfield  {author} {\bibinfo {author} {\bibfnamefont {M.~J.~H.}\
  \bibnamefont {Ku}}, \bibinfo {author} {\bibfnamefont {T.~X.}\ \bibnamefont
  {Zhou}}, \bibinfo {author} {\bibfnamefont {Q.}~\bibnamefont {Li}}, \bibinfo
  {author} {\bibfnamefont {Y.~J.}\ \bibnamefont {Shin}}, \bibinfo {author}
  {\bibfnamefont {J.~K.}\ \bibnamefont {Shi}}, \bibinfo {author} {\bibfnamefont
  {C.}~\bibnamefont {Burch}}, \bibinfo {author} {\bibfnamefont {L.~E.}\
  \bibnamefont {Anderson}}, \bibinfo {author} {\bibfnamefont {A.~T.}\
  \bibnamefont {Pierce}}, \bibinfo {author} {\bibfnamefont {Y.}~\bibnamefont
  {Xie}}, \bibinfo {author} {\bibfnamefont {A.}~\bibnamefont {Hamo}}, \bibinfo
  {author} {\bibfnamefont {U.}~\bibnamefont {Vool}}, \bibinfo {author}
  {\bibfnamefont {H.}~\bibnamefont {Zhang}}, \bibinfo {author} {\bibfnamefont
  {F.}~\bibnamefont {Casola}}, \bibinfo {author} {\bibfnamefont
  {T.}~\bibnamefont {Taniguchi}}, \bibinfo {author} {\bibfnamefont
  {K.}~\bibnamefont {Watanabe}}, \bibinfo {author} {\bibfnamefont {M.~M.}\
  \bibnamefont {Fogler}}, \bibinfo {author} {\bibfnamefont {P.}~\bibnamefont
  {Kim}}, \bibinfo {author} {\bibfnamefont {A.}~\bibnamefont {Yacoby}},\ and\
  \bibinfo {author} {\bibfnamefont {R.~L.}\ \bibnamefont {Walsworth}},\
  }\bibfield  {title} {\bibinfo {title} {Imaging viscous flow of the {{Dirac}}
  fluid in graphene.},\ }\href {https://doi.org/10.1038/s41586-020-2507-2}
  {\bibfield  {journal} {\bibinfo  {journal} {Nature}\ }\textbf {\bibinfo
  {volume} {583}},\ \bibinfo {pages} {537} (\bibinfo {year}
  {2020})}\BibitemShut {NoStop}%
\bibitem [{\citenamefont {Jenkins}\ \emph {et~al.}(2020)\citenamefont
  {Jenkins}, \citenamefont {Baumann}, \citenamefont {Zhou}, \citenamefont
  {Meynell}, \citenamefont {Yang}, \citenamefont {Watanabe}, \citenamefont
  {Taniguchi}, \citenamefont {Lucas}, \citenamefont {Young},\ and\
  \citenamefont {Jayich}}]{Jenkins2020}%
  \BibitemOpen
  \bibfield  {author} {\bibinfo {author} {\bibfnamefont {A.}~\bibnamefont
  {Jenkins}}, \bibinfo {author} {\bibfnamefont {S.}~\bibnamefont {Baumann}},
  \bibinfo {author} {\bibfnamefont {H.}~\bibnamefont {Zhou}}, \bibinfo {author}
  {\bibfnamefont {S.~A.}\ \bibnamefont {Meynell}}, \bibinfo {author}
  {\bibfnamefont {D.}~\bibnamefont {Yang}}, \bibinfo {author} {\bibfnamefont
  {K.}~\bibnamefont {Watanabe}}, \bibinfo {author} {\bibfnamefont
  {T.}~\bibnamefont {Taniguchi}}, \bibinfo {author} {\bibfnamefont
  {A.}~\bibnamefont {Lucas}}, \bibinfo {author} {\bibfnamefont {A.~F.}\
  \bibnamefont {Young}},\ and\ \bibinfo {author} {\bibfnamefont {A.~C.~B.}\
  \bibnamefont {Jayich}},\ }\href@noop {} {\bibinfo {title} {Imaging the
  breakdown of ohmic transport in graphene}} (\bibinfo {year} {2020}),\ \Eprint
  {https://arxiv.org/abs/2002.05065} {arXiv:2002.05065 [cond-mat.mes-hall]}
  \BibitemShut {NoStop}%
\bibitem [{\citenamefont {Han}\ \emph {et~al.}(2014)\citenamefont {Han},
  \citenamefont {Kawakami}, \citenamefont {Gmitra},\ and\ \citenamefont
  {Fabian}}]{Han2014}%
  \BibitemOpen
  \bibfield  {author} {\bibinfo {author} {\bibfnamefont {W.}~\bibnamefont
  {Han}}, \bibinfo {author} {\bibfnamefont {R.~K.}\ \bibnamefont {Kawakami}},
  \bibinfo {author} {\bibfnamefont {M.}~\bibnamefont {Gmitra}},\ and\ \bibinfo
  {author} {\bibfnamefont {J.}~\bibnamefont {Fabian}},\ }\bibfield  {title}
  {\bibinfo {title} {Graphene spintronics.},\ }\href
  {https://doi.org/10.1038/nnano.2014.214} {\bibfield  {journal} {\bibinfo
  {journal} {Nat. Nanotechnol.}\ }\textbf {\bibinfo {volume} {9}},\ \bibinfo
  {pages} {794} (\bibinfo {year} {2014})}\BibitemShut {NoStop}%
\bibitem [{\citenamefont {Schaibley}\ \emph {et~al.}(2016)\citenamefont
  {Schaibley}, \citenamefont {Yu}, \citenamefont {Clark}, \citenamefont
  {Rivera}, \citenamefont {Ross}, \citenamefont {Seyler}, \citenamefont {Yao},\
  and\ \citenamefont {Xu}}]{Schaibley2016}%
  \BibitemOpen
  \bibfield  {author} {\bibinfo {author} {\bibfnamefont {J.~R.}\ \bibnamefont
  {Schaibley}}, \bibinfo {author} {\bibfnamefont {H.}~\bibnamefont {Yu}},
  \bibinfo {author} {\bibfnamefont {G.}~\bibnamefont {Clark}}, \bibinfo
  {author} {\bibfnamefont {P.}~\bibnamefont {Rivera}}, \bibinfo {author}
  {\bibfnamefont {J.~S.}\ \bibnamefont {Ross}}, \bibinfo {author}
  {\bibfnamefont {K.~L.}\ \bibnamefont {Seyler}}, \bibinfo {author}
  {\bibfnamefont {W.}~\bibnamefont {Yao}},\ and\ \bibinfo {author}
  {\bibfnamefont {X.}~\bibnamefont {Xu}},\ }\bibfield  {title} {\bibinfo
  {title} {Valleytronics in {{2D}} materials},\ }\bibfield  {journal} {\bibinfo
   {journal} {Nat. Rev. Mater.}\ }\textbf {\bibinfo {volume} {1}},\ \href
  {https://doi.org/10.1038/natrevmats.2016.55} {10.1038/natrevmats.2016.55}
  (\bibinfo {year} {2016})\BibitemShut {NoStop}%
\bibitem [{\citenamefont {{\v Z}uti{\'c}}\ \emph {et~al.}(2004)\citenamefont
  {{\v Z}uti{\'c}}, \citenamefont {Fabian},\ and\ \citenamefont
  {Das~Sarma}}]{Zutic2004}%
  \BibitemOpen
  \bibfield  {author} {\bibinfo {author} {\bibfnamefont {I.}~\bibnamefont {{\v
  Z}uti{\'c}}}, \bibinfo {author} {\bibfnamefont {J.}~\bibnamefont {Fabian}},\
  and\ \bibinfo {author} {\bibfnamefont {S.}~\bibnamefont {Das~Sarma}},\
  }\bibfield  {title} {\bibinfo {title} {Spintronics: {{Fundamentals}} and
  applications},\ }\href@noop {} {\bibfield  {journal} {\bibinfo  {journal}
  {Rev. Mod. Phys.}\ }\textbf {\bibinfo {volume} {76}},\ \bibinfo {pages} {323}
  (\bibinfo {year} {2004})}\BibitemShut {NoStop}%
\bibitem [{\citenamefont {Avsar}\ \emph {et~al.}(2020)\citenamefont {Avsar},
  \citenamefont {Ochoa}, \citenamefont {Guinea}, \citenamefont {{\"O}zyilmaz},
  \citenamefont {{van Wees}},\ and\ \citenamefont {{Vera-Marun}}}]{Avsar2020}%
  \BibitemOpen
  \bibfield  {author} {\bibinfo {author} {\bibfnamefont {A.}~\bibnamefont
  {Avsar}}, \bibinfo {author} {\bibfnamefont {H.}~\bibnamefont {Ochoa}},
  \bibinfo {author} {\bibfnamefont {F.}~\bibnamefont {Guinea}}, \bibinfo
  {author} {\bibfnamefont {B.}~\bibnamefont {{\"O}zyilmaz}}, \bibinfo {author}
  {\bibfnamefont {B.~J.}\ \bibnamefont {{van Wees}}},\ and\ \bibinfo {author}
  {\bibfnamefont {I.~J.}\ \bibnamefont {{Vera-Marun}}},\ }\bibfield  {title}
  {\bibinfo {title} {Colloquium : {{Spintronics}} in graphene and other
  two-dimensional materials},\ }\href
  {https://doi.org/10.1103/revmodphys.92.021003} {\bibfield  {journal}
  {\bibinfo  {journal} {Rev. Mod. Phys.}\ }\textbf {\bibinfo {volume} {92}},\
  \bibinfo {pages} {021003} (\bibinfo {year} {2020})}\BibitemShut {NoStop}%
\bibitem [{\citenamefont {Mishchenko}\ \emph {et~al.}(2004)\citenamefont
  {Mishchenko}, \citenamefont {Shytov},\ and\ \citenamefont
  {Halperin}}]{Mishchenko2004}%
  \BibitemOpen
  \bibfield  {author} {\bibinfo {author} {\bibfnamefont {E.~G.}\ \bibnamefont
  {Mishchenko}}, \bibinfo {author} {\bibfnamefont {A.~V.}\ \bibnamefont
  {Shytov}},\ and\ \bibinfo {author} {\bibfnamefont {B.~I.}\ \bibnamefont
  {Halperin}},\ }\bibfield  {title} {\bibinfo {title} {Spin {{Current}} and
  {{Polarization}} in {{Impure Two}}-{{Dimensional Electron Systems}} with
  {{Spin}}-{{Orbit Coupling}}},\ }\href
  {https://doi.org/10.1103/physrevlett.93.226602} {\bibfield  {journal}
  {\bibinfo  {journal} {Phys. Rev. Lett.}\ }\textbf {\bibinfo {volume} {93}},\
  \bibinfo {pages} {226602} (\bibinfo {year} {2004})}\BibitemShut {NoStop}%
\bibitem [{\citenamefont {Flensberg}\ \emph {et~al.}(2001)\citenamefont
  {Flensberg}, \citenamefont {Jensen},\ and\ \citenamefont
  {Mortensen}}]{Flensberg2001}%
  \BibitemOpen
  \bibfield  {author} {\bibinfo {author} {\bibfnamefont {K.}~\bibnamefont
  {Flensberg}}, \bibinfo {author} {\bibfnamefont {T.~S.}\ \bibnamefont
  {Jensen}},\ and\ \bibinfo {author} {\bibfnamefont {N.~A.}\ \bibnamefont
  {Mortensen}},\ }\bibfield  {title} {\bibinfo {title} {Diffusion equation and
  spin drag in spin-polarized transport},\ }\href
  {https://doi.org/10.1103/physrevb.64.245308} {\bibfield  {journal} {\bibinfo
  {journal} {Phys. Rev. B}\ }\textbf {\bibinfo {volume} {64}},\ \bibinfo
  {pages} {245308} (\bibinfo {year} {2001})}\BibitemShut {NoStop}%
\bibitem [{\citenamefont {Weber}\ \emph {et~al.}(2005)\citenamefont {Weber},
  \citenamefont {Gedik}, \citenamefont {Moore}, \citenamefont {Orenstein},
  \citenamefont {Stephens},\ and\ \citenamefont {Awschalom}}]{Weber2005}%
  \BibitemOpen
  \bibfield  {author} {\bibinfo {author} {\bibfnamefont {C.~P.}\ \bibnamefont
  {Weber}}, \bibinfo {author} {\bibfnamefont {N.}~\bibnamefont {Gedik}},
  \bibinfo {author} {\bibfnamefont {J.~E.}\ \bibnamefont {Moore}}, \bibinfo
  {author} {\bibfnamefont {J.}~\bibnamefont {Orenstein}}, \bibinfo {author}
  {\bibfnamefont {J.}~\bibnamefont {Stephens}},\ and\ \bibinfo {author}
  {\bibfnamefont {D.~D.}\ \bibnamefont {Awschalom}},\ }\bibfield  {title}
  {\bibinfo {title} {Observation of spin {{Coulomb}} drag in a two-dimensional
  electron gas},\ }\href {https://doi.org/10.1038/nature04206} {\bibfield
  {journal} {\bibinfo  {journal} {Nature}\ }\textbf {\bibinfo {volume} {437}},\
  \bibinfo {pages} {1330} (\bibinfo {year} {2005})}\BibitemShut {NoStop}%
\bibitem [{\citenamefont {Leggett}\ and\ \citenamefont
  {Rice}(1968)}]{Leggett1968}%
  \BibitemOpen
  \bibfield  {author} {\bibinfo {author} {\bibfnamefont {A.~J.}\ \bibnamefont
  {Leggett}}\ and\ \bibinfo {author} {\bibfnamefont {M.~J.}\ \bibnamefont
  {Rice}},\ }\bibfield  {title} {\bibinfo {title} {Spin {{Echoes}} in {{Liquid
  He3}} and {{Mixtures}}: {{A Predicted New Effect}}},\ }\href
  {https://doi.org/10.1103/physrevlett.20.586} {\bibfield  {journal} {\bibinfo
  {journal} {Phys. Rev. Lett.}\ }\textbf {\bibinfo {volume} {20}},\ \bibinfo
  {pages} {586} (\bibinfo {year} {1968})}\BibitemShut {NoStop}%
\bibitem [{\citenamefont {Miyake}\ and\ \citenamefont
  {Mullin}(1983)}]{Miyake1983}%
  \BibitemOpen
  \bibfield  {author} {\bibinfo {author} {\bibfnamefont {K.}~\bibnamefont
  {Miyake}}\ and\ \bibinfo {author} {\bibfnamefont {W.~J.}\ \bibnamefont
  {Mullin}},\ }\bibfield  {title} {\bibinfo {title} {Spin {{Diffusion}} in a
  {{Two}}-{{Dimensional Degenerate Fermi Liquid}}},\ }\href
  {https://doi.org/10.1103/physrevlett.50.197} {\bibfield  {journal} {\bibinfo
  {journal} {Phys. Rev. Lett.}\ }\textbf {\bibinfo {volume} {50}},\ \bibinfo
  {pages} {197} (\bibinfo {year} {1983})}\BibitemShut {NoStop}%
\bibitem [{\citenamefont {M{\"u}ller}\ and\ \citenamefont
  {Nguyen}(2011)}]{Muller2011}%
  \BibitemOpen
  \bibfield  {author} {\bibinfo {author} {\bibfnamefont {M.}~\bibnamefont
  {M{\"u}ller}}\ and\ \bibinfo {author} {\bibfnamefont {H.~C.}\ \bibnamefont
  {Nguyen}},\ }\bibfield  {title} {\bibinfo {title} {Collision-dominated spin
  transport in graphene and {{Fermi}} liquids},\ }\href
  {https://doi.org/10.1088/1367-2630/13/3/035009} {\bibfield  {journal}
  {\bibinfo  {journal} {New J. Phys.}\ }\textbf {\bibinfo {volume} {13}},\
  \bibinfo {pages} {035009} (\bibinfo {year} {2011})}\BibitemShut {NoStop}%
\bibitem [{\citenamefont {{San-Jose}}\ \emph {et~al.}(2014)\citenamefont
  {{San-Jose}}, \citenamefont {{Guti{\'e}rrez-Rubio}}, \citenamefont {Sturla},\
  and\ \citenamefont {Guinea}}]{San-Jose2014}%
  \BibitemOpen
  \bibfield  {author} {\bibinfo {author} {\bibfnamefont {P.}~\bibnamefont
  {{San-Jose}}}, \bibinfo {author} {\bibfnamefont {A.}~\bibnamefont
  {{Guti{\'e}rrez-Rubio}}}, \bibinfo {author} {\bibfnamefont {M.}~\bibnamefont
  {Sturla}},\ and\ \bibinfo {author} {\bibfnamefont {F.}~\bibnamefont
  {Guinea}},\ }\bibfield  {title} {\bibinfo {title} {Spontaneous strains and
  gap in graphene on boron nitride},\ }\href
  {https://doi.org/10.1103/physrevb.90.075428} {\bibfield  {journal} {\bibinfo
  {journal} {Phys. Rev. B}\ }\textbf {\bibinfo {volume} {90}},\ \bibinfo
  {pages} {075428} (\bibinfo {year} {2014})}\BibitemShut {NoStop}%
\bibitem [{\citenamefont {Beenakker}(2008)}]{Beenakker2008}%
  \BibitemOpen
  \bibfield  {author} {\bibinfo {author} {\bibfnamefont {C.~W.~J.}\
  \bibnamefont {Beenakker}},\ }\bibfield  {title} {\bibinfo {title}
  {Colloquium: {{Andreev}} reflection and {{Klein}} tunneling in graphene},\
  }\href {https://doi.org/10.1103/revmodphys.80.1337} {\bibfield  {journal}
  {\bibinfo  {journal} {Rev. Mod. Phys.}\ }\textbf {\bibinfo {volume} {80}},\
  \bibinfo {pages} {1337} (\bibinfo {year} {2008})}\BibitemShut {NoStop}%
\bibitem [{\citenamefont {Hwang}\ and\ \citenamefont
  {Das~Sarma}(2008)}]{Hwang2008}%
  \BibitemOpen
  \bibfield  {author} {\bibinfo {author} {\bibfnamefont {E.~H.}\ \bibnamefont
  {Hwang}}\ and\ \bibinfo {author} {\bibfnamefont {S.}~\bibnamefont
  {Das~Sarma}},\ }\bibfield  {title} {\bibinfo {title} {Single-particle
  relaxation time versus transport scattering time in a two-dimensional
  graphene layer},\ }\href {https://doi.org/10.1103/physrevb.77.195412}
  {\bibfield  {journal} {\bibinfo  {journal} {Phys. Rev. B}\ }\textbf {\bibinfo
  {volume} {77}},\ \bibinfo {pages} {195412} (\bibinfo {year}
  {2008})}\BibitemShut {NoStop}%
\bibitem [{\citenamefont {Aleiner}\ \emph {et~al.}(2007)\citenamefont
  {Aleiner}, \citenamefont {Kharzeev},\ and\ \citenamefont
  {Tsvelik}}]{Aleiner2007}%
  \BibitemOpen
  \bibfield  {author} {\bibinfo {author} {\bibfnamefont {I.~L.}\ \bibnamefont
  {Aleiner}}, \bibinfo {author} {\bibfnamefont {D.~E.}\ \bibnamefont
  {Kharzeev}},\ and\ \bibinfo {author} {\bibfnamefont {A.~M.}\ \bibnamefont
  {Tsvelik}},\ }\bibfield  {title} {\bibinfo {title} {Spontaneous symmetry
  breaking in graphene subjected to an in-plane magnetic field},\ }\href
  {https://doi.org/10.1103/physrevb.76.195415} {\bibfield  {journal} {\bibinfo
  {journal} {Phys. Rev. B}\ }\textbf {\bibinfo {volume} {76}},\ \bibinfo
  {pages} {195415} (\bibinfo {year} {2007})}\BibitemShut {NoStop}%
\bibitem [{\citenamefont {Kharitonov}(2012)}]{Kharitonov2012}%
  \BibitemOpen
  \bibfield  {author} {\bibinfo {author} {\bibfnamefont {M.}~\bibnamefont
  {Kharitonov}},\ }\bibfield  {title} {\bibinfo {title} {Phase diagram for the
  {$N$}=0 quantum {{Hall}} state in monolayer graphene},\ }\href
  {https://doi.org/10.1103/physrevb.85.155439} {\bibfield  {journal} {\bibinfo
  {journal} {Phys. Rev. B}\ }\textbf {\bibinfo {volume} {85}},\ \bibinfo
  {pages} {155439} (\bibinfo {year} {2012})}\BibitemShut {NoStop}%
\bibitem [{Note1()}]{Note1}%
  \BibitemOpen
  \bibinfo {note} {Here we use $\Sigma $ for the sub-lattice Pauli matrices and
  reserve $\sigma $ for the spin Pauli matrices, as the sub-lattice degree of
  freedom will be projected out shortly. This differs from the notation
  commonly used in the literature of denoting the sub-lattice Pauli matrices
  $\sigma $ and spin Pauli matrices $s$. We retain the use of $\tau $ for
  valley Pauli matrices.}\BibitemShut {Stop}%
\bibitem [{\citenamefont {Nozieres}\ and\ \citenamefont
  {Pines}(1999)}]{Nozieres1999}%
  \BibitemOpen
  \bibfield  {author} {\bibinfo {author} {\bibfnamefont {P.}~\bibnamefont
  {Nozieres}}\ and\ \bibinfo {author} {\bibfnamefont {D.}~\bibnamefont
  {Pines}},\ }\href@noop {} {\emph {\bibinfo {title} {Theory {{Of Quantum
  Liquids}}}}},\ Advanced {{Books Classics}}\ (\bibinfo  {publisher} {{Avalon
  Publishing}},\ \bibinfo {year} {1999})\BibitemShut {NoStop}%
\bibitem [{\citenamefont {Xiao}\ \emph {et~al.}(2007)\citenamefont {Xiao},
  \citenamefont {Yao},\ and\ \citenamefont {Niu}}]{Xiao2007a}%
  \BibitemOpen
  \bibfield  {author} {\bibinfo {author} {\bibfnamefont {D.}~\bibnamefont
  {Xiao}}, \bibinfo {author} {\bibfnamefont {W.}~\bibnamefont {Yao}},\ and\
  \bibinfo {author} {\bibfnamefont {Q.}~\bibnamefont {Niu}},\ }\bibfield
  {title} {\bibinfo {title} {Valley-{{Contrasting Physics}} in {{Graphene}}:
  {{Magnetic Moment}} and {{Topological Transport}}},\ }\href
  {https://doi.org/10.1103/physrevlett.99.236809} {\bibfield  {journal}
  {\bibinfo  {journal} {Phys. Rev. Lett.}\ }\textbf {\bibinfo {volume} {99}},\
  \bibinfo {pages} {236809} (\bibinfo {year} {2007})}\BibitemShut {NoStop}%
\bibitem [{\citenamefont {Chen}\ and\ \citenamefont {Son}(2017)}]{Chen2017}%
  \BibitemOpen
  \bibfield  {author} {\bibinfo {author} {\bibfnamefont {J.-Y.}\ \bibnamefont
  {Chen}}\ and\ \bibinfo {author} {\bibfnamefont {D.~T.}\ \bibnamefont {Son}},\
  }\bibfield  {title} {\bibinfo {title} {Berry {{Fermi}} liquid theory},\
  }\href {https://doi.org/10.1016/j.aop.2016.12.017} {\bibfield  {journal}
  {\bibinfo  {journal} {Ann. Phys.}\ }\textbf {\bibinfo {volume} {377}},\
  \bibinfo {pages} {345} (\bibinfo {year} {2017})}\BibitemShut {NoStop}%
\bibitem [{\citenamefont {Stern}(1967)}]{Stern1967}%
  \BibitemOpen
  \bibfield  {author} {\bibinfo {author} {\bibfnamefont {F.}~\bibnamefont
  {Stern}},\ }\bibfield  {title} {\bibinfo {title} {Polarizability of a
  {{Two}}-{{Dimensional Electron Gas}}},\ }\href
  {https://doi.org/10.1103/physrevlett.18.546} {\bibfield  {journal} {\bibinfo
  {journal} {Phys. Rev. Lett.}\ }\textbf {\bibinfo {volume} {18}},\ \bibinfo
  {pages} {546} (\bibinfo {year} {1967})}\BibitemShut {NoStop}%
\bibitem [{\citenamefont {Hwang}\ and\ \citenamefont
  {Das~Sarma}(2007)}]{Hwang2007}%
  \BibitemOpen
  \bibfield  {author} {\bibinfo {author} {\bibfnamefont {E.~H.}\ \bibnamefont
  {Hwang}}\ and\ \bibinfo {author} {\bibfnamefont {S.}~\bibnamefont
  {Das~Sarma}},\ }\bibfield  {title} {\bibinfo {title} {Dielectric function,
  screening, and plasmons in two-dimensional graphene},\ }\href
  {https://doi.org/10.1103/physrevb.75.205418} {\bibfield  {journal} {\bibinfo
  {journal} {Phys. Rev. B}\ }\textbf {\bibinfo {volume} {75}},\ \bibinfo
  {pages} {205418} (\bibinfo {year} {2007})}\BibitemShut {NoStop}%
\bibitem [{\citenamefont {Lucas}\ and\ \citenamefont
  {Das~Sarma}(2018)}]{Lucas2018}%
  \BibitemOpen
  \bibfield  {author} {\bibinfo {author} {\bibfnamefont {A.}~\bibnamefont
  {Lucas}}\ and\ \bibinfo {author} {\bibfnamefont {S.}~\bibnamefont
  {Das~Sarma}},\ }\bibfield  {title} {\bibinfo {title} {Electronic sound modes
  and plasmons in hydrodynamic two-dimensional metals},\ }\href
  {https://doi.org/10.1103/physrevb.97.115449} {\bibfield  {journal} {\bibinfo
  {journal} {Phys. Rev. B}\ }\textbf {\bibinfo {volume} {97}},\ \bibinfo
  {pages} {115449} (\bibinfo {year} {2018})}\BibitemShut {NoStop}%
\bibitem [{\citenamefont {Lif{\v s}ic}\ \emph {et~al.}(2006)\citenamefont
  {Lif{\v s}ic}, \citenamefont {Pitaevskij}, \citenamefont {Landau},\ and\
  \citenamefont {Lifshitz}}]{Lifsic2006}%
  \BibitemOpen
  \bibfield  {author} {\bibinfo {author} {\bibfnamefont {E.~M.}\ \bibnamefont
  {Lif{\v s}ic}}, \bibinfo {author} {\bibfnamefont {L.~P.}\ \bibnamefont
  {Pitaevskij}}, \bibinfo {author} {\bibfnamefont {L.~D.}\ \bibnamefont
  {Landau}},\ and\ \bibinfo {author} {\bibfnamefont {E.~M.}\ \bibnamefont
  {Lifshitz}},\ }\href@noop {} {\emph {\bibinfo {title} {Statistical Physics.
  {{Part}} 2. {{Theory}} of the Condensed State}}},\ \bibinfo {edition}
  {reprinted}\ ed.,\ \bibinfo {series} {Course of Theoretical Physics}\ No.\
  \bibinfo {number} {by E. M. Lifshitz and L. P. Pitaevski{\u \i}; Vol.
  9[...]}\ (\bibinfo  {publisher} {{Elsevier}},\ \bibinfo {address}
  {{Oxford}},\ \bibinfo {year} {2006})\ p.\ \bibinfo {pages} {387}\BibitemShut
  {NoStop}%
\bibitem [{\citenamefont {Klein}\ \emph {et~al.}(2019)\citenamefont {Klein},
  \citenamefont {Maslov}, \citenamefont {Pitaevskii},\ and\ \citenamefont
  {Chubukov}}]{Klein2019}%
  \BibitemOpen
  \bibfield  {author} {\bibinfo {author} {\bibfnamefont {A.}~\bibnamefont
  {Klein}}, \bibinfo {author} {\bibfnamefont {D.~L.}\ \bibnamefont {Maslov}},
  \bibinfo {author} {\bibfnamefont {L.~P.}\ \bibnamefont {Pitaevskii}},\ and\
  \bibinfo {author} {\bibfnamefont {A.~V.}\ \bibnamefont {Chubukov}},\
  }\bibfield  {title} {\bibinfo {title} {Collective modes near a
  {{Pomeranchuk}} instability in two dimensions},\ }\href
  {https://doi.org/10.1103/physrevresearch.1.033134} {\bibfield  {journal}
  {\bibinfo  {journal} {Phys. Rev. Res.}\ }\textbf {\bibinfo {volume} {1}},\
  \bibinfo {pages} {033134} (\bibinfo {year} {2019})}\BibitemShut {NoStop}%
\bibitem [{\citenamefont {Klein}\ \emph {et~al.}(2020)\citenamefont {Klein},
  \citenamefont {Maslov},\ and\ \citenamefont {Chubukov}}]{Klein2020}%
  \BibitemOpen
  \bibfield  {author} {\bibinfo {author} {\bibfnamefont {A.}~\bibnamefont
  {Klein}}, \bibinfo {author} {\bibfnamefont {D.~L.}\ \bibnamefont {Maslov}},\
  and\ \bibinfo {author} {\bibfnamefont {A.~V.}\ \bibnamefont {Chubukov}},\
  }\bibfield  {title} {\bibinfo {title} {Hidden and mirage collective modes in
  two dimensional {{Fermi}} liquids},\ }\href
  {https://doi.org/10.1038/s41535-020-0250-4} {\bibfield  {journal} {\bibinfo
  {journal} {npj Quantum Materials}\ }\textbf {\bibinfo {volume} {5}},\
  \bibinfo {pages} {55} (\bibinfo {year} {2020})}\BibitemShut {NoStop}%
\bibitem [{\citenamefont {Hwang}\ and\ \citenamefont
  {Das~Sarma}(2009)}]{Hwang2009}%
  \BibitemOpen
  \bibfield  {author} {\bibinfo {author} {\bibfnamefont {E.~H.}\ \bibnamefont
  {Hwang}}\ and\ \bibinfo {author} {\bibfnamefont {S.}~\bibnamefont
  {Das~Sarma}},\ }\bibfield  {title} {\bibinfo {title} {Screening-induced
  temperature-dependent transport in two-dimensional graphene},\ }\href
  {https://doi.org/10.1103/physrevb.79.165404} {\bibfield  {journal} {\bibinfo
  {journal} {Phys. Rev. B}\ }\textbf {\bibinfo {volume} {79}},\ \bibinfo
  {pages} {165404} (\bibinfo {year} {2009})}\BibitemShut {NoStop}%
\bibitem [{\citenamefont {Ledwith}\ \emph
  {et~al.}(2019{\natexlab{a}})\citenamefont {Ledwith}, \citenamefont {Guo},\
  and\ \citenamefont {Levitov}}]{Ledwith2019}%
  \BibitemOpen
  \bibfield  {author} {\bibinfo {author} {\bibfnamefont {P.~J.}\ \bibnamefont
  {Ledwith}}, \bibinfo {author} {\bibfnamefont {H.}~\bibnamefont {Guo}},\ and\
  \bibinfo {author} {\bibfnamefont {L.}~\bibnamefont {Levitov}},\ }\bibfield
  {title} {\bibinfo {title} {The hierarchy of excitation lifetimes in
  two-dimensional {{Fermi}} gases},\ }\href
  {https://doi.org/10.1016/j.aop.2019.167913} {\bibfield  {journal} {\bibinfo
  {journal} {Ann. Phys.}\ }\textbf {\bibinfo {volume} {411}},\ \bibinfo {pages}
  {167913} (\bibinfo {year} {2019}{\natexlab{a}})}\BibitemShut {NoStop}%
\bibitem [{\citenamefont {Ledwith}\ \emph
  {et~al.}(2019{\natexlab{b}})\citenamefont {Ledwith}, \citenamefont {Guo},\
  and\ \citenamefont {Levitov}}]{Ledwith2017}%
  \BibitemOpen
  \bibfield  {author} {\bibinfo {author} {\bibfnamefont {P.}~\bibnamefont
  {Ledwith}}, \bibinfo {author} {\bibfnamefont {H.}~\bibnamefont {Guo}},\ and\
  \bibinfo {author} {\bibfnamefont {L.}~\bibnamefont {Levitov}},\ }\href@noop
  {} {\bibinfo {title} {Angular {{Superdiffusion}} and {{Directional Memory}}
  in {{Two}}-{{Dimensional Electron Fluids}}}} (\bibinfo {year}
  {2019}{\natexlab{b}}),\ \Eprint {https://arxiv.org/abs/1708.01915}
  {arXiv:1708.01915 [cond-mat.mes-hall]} \BibitemShut {NoStop}%
\bibitem [{Note2()}]{Note2}%
  \BibitemOpen
  \bibinfo {note} {Because of the time-reversal symmetry of the system this is
  equivalent to calculating $\tau ^\mu _\protect \text {tr} \equiv \tau ^\mu
  _1$.}\BibitemShut {Stop}%
\bibitem [{\citenamefont {Menashe}\ and\ \citenamefont
  {Laikhtman}(1996)}]{Menashe1996}%
  \BibitemOpen
  \bibfield  {author} {\bibinfo {author} {\bibfnamefont {D.}~\bibnamefont
  {Menashe}}\ and\ \bibinfo {author} {\bibfnamefont {B.}~\bibnamefont
  {Laikhtman}},\ }\bibfield  {title} {\bibinfo {title} {Quasiparticle lifetime
  in a two-dimensional electron system in the limit of low temperature and
  excitation energy},\ }\href {https://doi.org/10.1103/physrevb.54.11561}
  {\bibfield  {journal} {\bibinfo  {journal} {Phys. Rev. B}\ }\textbf {\bibinfo
  {volume} {54}},\ \bibinfo {pages} {11561} (\bibinfo {year}
  {1996})}\BibitemShut {NoStop}%
\bibitem [{Note3()}]{Note3}%
  \BibitemOpen
  \bibinfo {note} {We perform calculations in the static screening
  approximation. Dynamic screening effects are unimportant so long as we remain
  at low temperatures $v q_{TF} \gg T$ (see \protect \cref {sec:forward}), but
  see Ref.~\protect \rev@citealp {Zala2001,*Narozhny2002} and the supplement to
  Ref.~\protect \rev@citealp {Alekseev2020} for discussion of the effects of
  dynamic screening.}\BibitemShut {Stop}%
\bibitem [{\citenamefont {Zala}\ \emph {et~al.}(2001)\citenamefont {Zala},
  \citenamefont {Narozhny},\ and\ \citenamefont {Aleiner}}]{Zala2001}%
  \BibitemOpen
  \bibfield  {author} {\bibinfo {author} {\bibfnamefont {G.}~\bibnamefont
  {Zala}}, \bibinfo {author} {\bibfnamefont {B.~N.}\ \bibnamefont {Narozhny}},\
  and\ \bibinfo {author} {\bibfnamefont {I.~L.}\ \bibnamefont {Aleiner}},\
  }\bibfield  {title} {\bibinfo {title} {Interaction corrections at
  intermediate temperatures: {{Longitudinal}} conductivity and kinetic
  equation},\ }\href {https://doi.org/10.1103/PhysRevB.64.214204} {\bibfield
  {journal} {\bibinfo  {journal} {Phys. Rev. B}\ }\textbf {\bibinfo {volume}
  {64}},\ \bibinfo {pages} {214204} (\bibinfo {year} {2001})}\BibitemShut
  {NoStop}%
\bibitem [{\citenamefont {Narozhny}\ \emph {et~al.}(2002)\citenamefont
  {Narozhny}, \citenamefont {Zala},\ and\ \citenamefont
  {Aleiner}}]{Narozhny2002}%
  \BibitemOpen
  \bibfield  {author} {\bibinfo {author} {\bibfnamefont {B.~N.}\ \bibnamefont
  {Narozhny}}, \bibinfo {author} {\bibfnamefont {G.}~\bibnamefont {Zala}},\
  and\ \bibinfo {author} {\bibfnamefont {I.~L.}\ \bibnamefont {Aleiner}},\
  }\bibfield  {title} {\bibinfo {title} {Interaction corrections at
  intermediate temperatures: {{Dephasing}} time},\ }\href
  {https://doi.org/10.1103/PhysRevB.65.180202} {\bibfield  {journal} {\bibinfo
  {journal} {Phys. Rev. B}\ }\textbf {\bibinfo {volume} {65}},\ \bibinfo
  {pages} {180202(R)} (\bibinfo {year} {2002})}\BibitemShut {NoStop}%
\bibitem [{\citenamefont {Alekseev}\ and\ \citenamefont
  {Dmitriev}(2020)}]{Alekseev2020}%
  \BibitemOpen
  \bibfield  {author} {\bibinfo {author} {\bibfnamefont {P.~S.}\ \bibnamefont
  {Alekseev}}\ and\ \bibinfo {author} {\bibfnamefont {A.~P.}\ \bibnamefont
  {Dmitriev}},\ }\bibfield  {title} {\bibinfo {title} {Viscosity of
  two-dimensional electrons},\ }\href
  {https://doi.org/10.1103/PhysRevB.102.241409} {\bibfield  {journal} {\bibinfo
   {journal} {Phys. Rev. B}\ }\textbf {\bibinfo {volume} {102}},\ \bibinfo
  {pages} {241409(R)} (\bibinfo {year} {2020})}\BibitemShut {NoStop}%
\bibitem [{\citenamefont {Abanin}\ \emph {et~al.}(2009)\citenamefont {Abanin},
  \citenamefont {Shytov}, \citenamefont {Levitov},\ and\ \citenamefont
  {Halperin}}]{Abanin2009}%
  \BibitemOpen
  \bibfield  {author} {\bibinfo {author} {\bibfnamefont {D.~A.}\ \bibnamefont
  {Abanin}}, \bibinfo {author} {\bibfnamefont {A.~V.}\ \bibnamefont {Shytov}},
  \bibinfo {author} {\bibfnamefont {L.~S.}\ \bibnamefont {Levitov}},\ and\
  \bibinfo {author} {\bibfnamefont {B.~I.}\ \bibnamefont {Halperin}},\
  }\bibfield  {title} {\bibinfo {title} {Nonlocal charge transport mediated by
  spin diffusion in the spin {{Hall}} effect regime},\ }\href
  {https://doi.org/10.1103/physrevb.79.035304} {\bibfield  {journal} {\bibinfo
  {journal} {Phys. Rev. B}\ }\textbf {\bibinfo {volume} {79}},\ \bibinfo
  {pages} {035304} (\bibinfo {year} {2009})}\BibitemShut {NoStop}%
\bibitem [{\citenamefont {Yamamoto}\ \emph {et~al.}(2015)\citenamefont
  {Yamamoto}, \citenamefont {Shimazaki}, \citenamefont {Borzenets},\ and\
  \citenamefont {Tarucha}}]{Yamamoto2015}%
  \BibitemOpen
  \bibfield  {author} {\bibinfo {author} {\bibfnamefont {M.}~\bibnamefont
  {Yamamoto}}, \bibinfo {author} {\bibfnamefont {Y.}~\bibnamefont {Shimazaki}},
  \bibinfo {author} {\bibfnamefont {I.~V.}\ \bibnamefont {Borzenets}},\ and\
  \bibinfo {author} {\bibfnamefont {S.}~\bibnamefont {Tarucha}},\ }\bibfield
  {title} {\bibinfo {title} {Valley {{Hall Effect}} in {{Two}}-{{Dimensional
  Hexagonal Lattices}}},\ }\href {https://doi.org/10.7566/jpsj.84.121006}
  {\bibfield  {journal} {\bibinfo  {journal} {J. Phys. Soc. Jpn.}\ }\textbf
  {\bibinfo {volume} {84}},\ \bibinfo {pages} {121006} (\bibinfo {year}
  {2015})}\BibitemShut {NoStop}%
\bibitem [{\citenamefont {Gorbachev}\ \emph {et~al.}(2014)\citenamefont
  {Gorbachev}, \citenamefont {Song}, \citenamefont {Yu}, \citenamefont
  {Kretinin}, \citenamefont {Withers}, \citenamefont {Cao}, \citenamefont
  {Mishchenko}, \citenamefont {Grigorieva}, \citenamefont {Novoselov},
  \citenamefont {Levitov},\ and\ \citenamefont {Geim}}]{Gorbachev2014}%
  \BibitemOpen
  \bibfield  {author} {\bibinfo {author} {\bibfnamefont {R.~V.}\ \bibnamefont
  {Gorbachev}}, \bibinfo {author} {\bibfnamefont {J.~C.~W.}\ \bibnamefont
  {Song}}, \bibinfo {author} {\bibfnamefont {G.~L.}\ \bibnamefont {Yu}},
  \bibinfo {author} {\bibfnamefont {A.~V.}\ \bibnamefont {Kretinin}}, \bibinfo
  {author} {\bibfnamefont {F.}~\bibnamefont {Withers}}, \bibinfo {author}
  {\bibfnamefont {Y.}~\bibnamefont {Cao}}, \bibinfo {author} {\bibfnamefont
  {A.}~\bibnamefont {Mishchenko}}, \bibinfo {author} {\bibfnamefont {I.~V.}\
  \bibnamefont {Grigorieva}}, \bibinfo {author} {\bibfnamefont {K.~S.}\
  \bibnamefont {Novoselov}}, \bibinfo {author} {\bibfnamefont {L.~S.}\
  \bibnamefont {Levitov}},\ and\ \bibinfo {author} {\bibfnamefont {A.~K.}\
  \bibnamefont {Geim}},\ }\bibfield  {title} {\bibinfo {title} {Detecting
  topological currents in graphene superlattices},\ }\href
  {https://doi.org/10.1126/science.1254966} {\bibfield  {journal} {\bibinfo
  {journal} {Science}\ }\textbf {\bibinfo {volume} {346}},\ \bibinfo {pages}
  {448} (\bibinfo {year} {2014})}\BibitemShut {NoStop}%
\bibitem [{\citenamefont {{Ingla-Ayn{\'e}s}}\ \emph {et~al.}(2015)\citenamefont
  {{Ingla-Ayn{\'e}s}}, \citenamefont {Guimar{\~a}es}, \citenamefont
  {Meijerink}, \citenamefont {Zomer},\ and\ \citenamefont {{van
  Wees}}}]{Ingla-Aynes2015}%
  \BibitemOpen
  \bibfield  {author} {\bibinfo {author} {\bibfnamefont {J.}~\bibnamefont
  {{Ingla-Ayn{\'e}s}}}, \bibinfo {author} {\bibfnamefont {M.~H.~D.}\
  \bibnamefont {Guimar{\~a}es}}, \bibinfo {author} {\bibfnamefont {R.~J.}\
  \bibnamefont {Meijerink}}, \bibinfo {author} {\bibfnamefont {P.~J.}\
  \bibnamefont {Zomer}},\ and\ \bibinfo {author} {\bibfnamefont {B.~J.}\
  \bibnamefont {{van Wees}}},\ }\bibfield  {title} {\bibinfo {title}
  {24-{$\mu$}m spin relaxation length in boron nitride encapsulated bilayer
  graphene},\ }\href {https://doi.org/10.1103/physrevb.92.201410} {\bibfield
  {journal} {\bibinfo  {journal} {Phys. Rev. B}\ }\textbf {\bibinfo {volume}
  {92}},\ \bibinfo {pages} {201410(R)} (\bibinfo {year} {2015})}\BibitemShut
  {NoStop}%
\bibitem [{\citenamefont {Gurram}\ \emph {et~al.}(2018)\citenamefont {Gurram},
  \citenamefont {Omar},\ and\ \citenamefont {{van Wees}}}]{Gurram2018}%
  \BibitemOpen
  \bibfield  {author} {\bibinfo {author} {\bibfnamefont {M.}~\bibnamefont
  {Gurram}}, \bibinfo {author} {\bibfnamefont {S.}~\bibnamefont {Omar}},\ and\
  \bibinfo {author} {\bibfnamefont {B.~J.}\ \bibnamefont {{van Wees}}},\
  }\bibfield  {title} {\bibinfo {title} {Electrical spin injection, transport,
  and detection in graphene-hexagonal boron nitride van der {{Waals}}
  heterostructures: Progress and perspectives},\ }\href
  {https://doi.org/10.1088/2053-1583/aac34d} {\bibfield  {journal} {\bibinfo
  {journal} {2D Mater.}\ }\textbf {\bibinfo {volume} {5}},\ \bibinfo {pages}
  {032004} (\bibinfo {year} {2018})}\BibitemShut {NoStop}%
\bibitem [{\citenamefont {Liu}\ and\ \citenamefont
  {Finkel'stein}(2020)}]{Liu2020}%
  \BibitemOpen
  \bibfield  {author} {\bibinfo {author} {\bibfnamefont {A.}~\bibnamefont
  {Liu}}\ and\ \bibinfo {author} {\bibfnamefont {A.~M.}\ \bibnamefont
  {Finkel'stein}},\ }\bibfield  {title} {\bibinfo {title} {Valley current in
  graphene through electron-phonon interaction},\ }\href
  {https://doi.org/10.1103/physrevb.101.241401} {\bibfield  {journal} {\bibinfo
   {journal} {Phys. Rev. B}\ }\textbf {\bibinfo {volume} {101}},\ \bibinfo
  {pages} {241401(R)} (\bibinfo {year} {2020})}\BibitemShut {NoStop}%
\bibitem [{\citenamefont {Lee}\ \emph {et~al.}(2016)\citenamefont {Lee},
  \citenamefont {Wallbank}, \citenamefont {Gallagher}, \citenamefont
  {Watanabe}, \citenamefont {Taniguchi}, \citenamefont {Fal'ko},\ and\
  \citenamefont {{Goldhaber-Gordon}}}]{Lee2016}%
  \BibitemOpen
  \bibfield  {author} {\bibinfo {author} {\bibfnamefont {M.}~\bibnamefont
  {Lee}}, \bibinfo {author} {\bibfnamefont {J.~R.}\ \bibnamefont {Wallbank}},
  \bibinfo {author} {\bibfnamefont {P.}~\bibnamefont {Gallagher}}, \bibinfo
  {author} {\bibfnamefont {K.}~\bibnamefont {Watanabe}}, \bibinfo {author}
  {\bibfnamefont {T.}~\bibnamefont {Taniguchi}}, \bibinfo {author}
  {\bibfnamefont {V.~I.}\ \bibnamefont {Fal'ko}},\ and\ \bibinfo {author}
  {\bibfnamefont {D.}~\bibnamefont {{Goldhaber-Gordon}}},\ }\bibfield  {title}
  {\bibinfo {title} {Ballistic miniband conduction in a graphene
  superlattice.},\ }\href {https://doi.org/10.1126/science.aaf1095} {\bibfield
  {journal} {\bibinfo  {journal} {Science}\ }\textbf {\bibinfo {volume}
  {353}},\ \bibinfo {pages} {1526} (\bibinfo {year} {2016})}\BibitemShut
  {NoStop}%
\bibitem [{\citenamefont {Glazov}\ and\ \citenamefont
  {Ivchenko}(2002)}]{Glazov2002}%
  \BibitemOpen
  \bibfield  {author} {\bibinfo {author} {\bibfnamefont {M.~M.}\ \bibnamefont
  {Glazov}}\ and\ \bibinfo {author} {\bibfnamefont {E.~L.}\ \bibnamefont
  {Ivchenko}},\ }\bibfield  {title} {\bibinfo {title} {Precession spin
  relaxation mechanism caused by frequent electron-electron collisions},\
  }\href {https://doi.org/10.1134/1.1490009} {\bibfield  {journal} {\bibinfo
  {journal} {JETP Lett.}\ }\textbf {\bibinfo {volume} {75}},\ \bibinfo {pages}
  {403} (\bibinfo {year} {2002})}\BibitemShut {NoStop}%
\bibitem [{\citenamefont {Glazov}\ and\ \citenamefont
  {Ivchenko}(2004)}]{Glazov2004}%
  \BibitemOpen
  \bibfield  {author} {\bibinfo {author} {\bibfnamefont {M.~M.}\ \bibnamefont
  {Glazov}}\ and\ \bibinfo {author} {\bibfnamefont {E.~L.}\ \bibnamefont
  {Ivchenko}},\ }\bibfield  {title} {\bibinfo {title} {Effect of
  electron-electron interaction on spin relaxation of charge carriers in
  semiconductors},\ }\href {https://doi.org/10.1134/1.1854815} {\bibfield
  {journal} {\bibinfo  {journal} {J. Exp. Theor. Phys.}\ }\textbf {\bibinfo
  {volume} {99}},\ \bibinfo {pages} {1279} (\bibinfo {year}
  {2004})}\BibitemShut {NoStop}%
\bibitem [{\citenamefont {Wang}\ \emph {et~al.}(2016)\citenamefont {Wang},
  \citenamefont {Ki}, \citenamefont {Khoo}, \citenamefont {Mauro},
  \citenamefont {Berger}, \citenamefont {Levitov},\ and\ \citenamefont
  {Morpurgo}}]{Wang2016c}%
  \BibitemOpen
  \bibfield  {author} {\bibinfo {author} {\bibfnamefont {Z.}~\bibnamefont
  {Wang}}, \bibinfo {author} {\bibfnamefont {D.-K.}\ \bibnamefont {Ki}},
  \bibinfo {author} {\bibfnamefont {J.~Y.}\ \bibnamefont {Khoo}}, \bibinfo
  {author} {\bibfnamefont {D.}~\bibnamefont {Mauro}}, \bibinfo {author}
  {\bibfnamefont {H.}~\bibnamefont {Berger}}, \bibinfo {author} {\bibfnamefont
  {L.~S.}\ \bibnamefont {Levitov}},\ and\ \bibinfo {author} {\bibfnamefont
  {A.~F.}\ \bibnamefont {Morpurgo}},\ }\bibfield  {title} {\bibinfo {title}
  {Origin and {{Magnitude}} of `{{Designer}}' {{Spin}}-{{Orbit Interaction}} in
  {{Graphene}} on {{Semiconducting Transition Metal Dichalcogenides}}},\ }\href
  {https://doi.org/10.1103/physrevx.6.041020} {\bibfield  {journal} {\bibinfo
  {journal} {Phys. Rev. X}\ }\textbf {\bibinfo {volume} {6}},\ \bibinfo {pages}
  {041020} (\bibinfo {year} {2016})}\BibitemShut {NoStop}%
\bibitem [{\citenamefont {Tarnopolsky}\ \emph {et~al.}(2019)\citenamefont
  {Tarnopolsky}, \citenamefont {Kruchkov},\ and\ \citenamefont
  {Vishwanath}}]{Tarnopolsky2019}%
  \BibitemOpen
  \bibfield  {author} {\bibinfo {author} {\bibfnamefont {G.}~\bibnamefont
  {Tarnopolsky}}, \bibinfo {author} {\bibfnamefont {A.~J.}\ \bibnamefont
  {Kruchkov}},\ and\ \bibinfo {author} {\bibfnamefont {A.}~\bibnamefont
  {Vishwanath}},\ }\bibfield  {title} {\bibinfo {title} {Origin of {{Magic
  Angles}} in {{Twisted Bilayer Graphene}}},\ }\href
  {https://doi.org/10.1103/physrevlett.122.106405} {\bibfield  {journal}
  {\bibinfo  {journal} {Phys. Rev. Lett.}\ }\textbf {\bibinfo {volume} {122}},\
  \bibinfo {pages} {106405} (\bibinfo {year} {2019})}\BibitemShut {NoStop}%
\bibitem [{\citenamefont {Vafek}\ and\ \citenamefont {Kang}(2020)}]{Vafek2020}%
  \BibitemOpen
  \bibfield  {author} {\bibinfo {author} {\bibfnamefont {O.}~\bibnamefont
  {Vafek}}\ and\ \bibinfo {author} {\bibfnamefont {J.}~\bibnamefont {Kang}},\
  }\bibfield  {title} {\bibinfo {title} {Renormalization {{Group Study}} of
  {{Hidden Symmetry}} in {{Twisted Bilayer Graphene}} with {{Coulomb
  Interactions}}},\ }\href {https://doi.org/10.1103/PhysRevLett.125.257602}
  {\bibfield  {journal} {\bibinfo  {journal} {Phys. Rev. Lett.}\ }\textbf
  {\bibinfo {volume} {125}},\ \bibinfo {pages} {257602} (\bibinfo {year}
  {2020})}\BibitemShut {NoStop}%
\bibitem [{\citenamefont {Emery}(1968)}]{Emery1968}%
  \BibitemOpen
  \bibfield  {author} {\bibinfo {author} {\bibfnamefont {V.~J.}\ \bibnamefont
  {Emery}},\ }\bibfield  {title} {\bibinfo {title} {Low-{{Temperature
  Expansion}} of the {{Transport Coefficients}} and {{Specific Heat}} of
  {{Fermi Liquids}}},\ }\href {https://doi.org/10.1103/physrev.170.205}
  {\bibfield  {journal} {\bibinfo  {journal} {Phys. Rev.}\ }\textbf {\bibinfo
  {volume} {170}},\ \bibinfo {pages} {205} (\bibinfo {year}
  {1968})}\BibitemShut {NoStop}%
\bibitem [{\citenamefont {Laikhtman}(1992)}]{Laikhtman1992}%
  \BibitemOpen
  \bibfield  {author} {\bibinfo {author} {\bibfnamefont {B.}~\bibnamefont
  {Laikhtman}},\ }\bibfield  {title} {\bibinfo {title} {Electron-electron
  angular relaxation in a two-dimensional electron gas},\ }\href
  {https://doi.org/10.1103/PhysRevB.45.1259} {\bibfield  {journal} {\bibinfo
  {journal} {Phys. Rev. B}\ }\textbf {\bibinfo {volume} {45}},\ \bibinfo
  {pages} {1259} (\bibinfo {year} {1992})}\BibitemShut {NoStop}%
\bibitem [{\citenamefont {Das~Sarma}\ and\ \citenamefont
  {Stern}(1985)}]{DasSarma1985}%
  \BibitemOpen
  \bibfield  {author} {\bibinfo {author} {\bibfnamefont {S.}~\bibnamefont
  {Das~Sarma}}\ and\ \bibinfo {author} {\bibfnamefont {F.}~\bibnamefont
  {Stern}},\ }\bibfield  {title} {\bibinfo {title} {Single-particle relaxation
  time versus scattering time in an impure electron gas},\ }\href
  {https://doi.org/10.1103/physrevb.32.8442} {\bibfield  {journal} {\bibinfo
  {journal} {Phys. Rev. B}\ }\textbf {\bibinfo {volume} {32}},\ \bibinfo
  {pages} {8442} (\bibinfo {year} {1985})}\BibitemShut {NoStop}%
\bibitem [{\citenamefont {M{\"u}ller}\ \emph {et~al.}(2009)\citenamefont
  {M{\"u}ller}, \citenamefont {Schmalian},\ and\ \citenamefont
  {Fritz}}]{Muller2009}%
  \BibitemOpen
  \bibfield  {author} {\bibinfo {author} {\bibfnamefont {M.}~\bibnamefont
  {M{\"u}ller}}, \bibinfo {author} {\bibfnamefont {J.}~\bibnamefont
  {Schmalian}},\ and\ \bibinfo {author} {\bibfnamefont {L.}~\bibnamefont
  {Fritz}},\ }\bibfield  {title} {\bibinfo {title} {Graphene: A nearly perfect
  fluid},\ }\href {https://doi.org/10.1103/PhysRevLett.103.025301} {\bibfield
  {journal} {\bibinfo  {journal} {Phys. Rev. Lett.}\ }\textbf {\bibinfo
  {volume} {103}},\ \bibinfo {pages} {025301} (\bibinfo {year}
  {2009})}\BibitemShut {NoStop}%
\end{thebibliography}%

%%%VVVVVVV
%% arXiv only
%% Nicely integrate supplemental material into the appendix
%%%%%%%
% \onecolumngrid
\appendix
% \renewcommand{\appendixpagename}{\centering Supplemental Material}
% \setcounter{secnumdepth}{2}
% \setcounter{section}{0}
% \setcounter{equation}{0}
% \setcounter{figure}{0}
% \setcounter{table}{0}
% \renewcommand{\thesection}{\Alph{section}}
% \renewcommand{\thesubsection}{\thesection.\arabic{subsection}}
% \appendixpage
%%%%
%^^^^^End arXiv only
%%%%%%

%% Number equations for supplemental material
% \renewcommand{\theequation}{S\arabic{equation}}
% \renewcommand{\thefigure}{S\arabic{figure}}
% \renewcommand{\thetable}{S\arabic{table}}

\section{Symmetry imposed form of the interaction functions}
\label{sec:sym-int}

In this section we derive the short ranged interaction terms which are consistent with the symmetry of the gapped graphene lattice and relate them to the interaction functions in the spin-valley channels introduced in \cref{eq:gint}.

\subsection{Symmetry Analysis of short ranged interactions}
Ignoring out of plane behavior the point group of gapped graphene is $C_{3v}$.
In the absence of spin-orbit coupling, which is generally weak in graphene, the local symmetry is enlarged to $C_{3V} \oplus SU(2)$.
In the valley-sublattice basis with spinor
\begin{equation}
  \Psi_\mathbf{k} =
  \begin{pmatrix}
   \psi_{KA}(\mathbf{k}) \\
   \psi_{KB}(\mathbf{k}) \\
   \psi_{K'B}(\mathbf{k}) \\
   - \psi_{K'A}(\mathbf{k}) \\
  \end{pmatrix}
\end{equation}
we can define the representative symmetry operators
\begin{equation}
 C_3: e^{i\frac{\pi}{3} \hat{\Sigma}_3},\quad \sigma_V: \hat{\Sigma}_2\hat{\tau}_2, \quad\Theta: \sigma_V K, \quad T: e^{\frac{2\pi i}{3}\tau^z}
\end{equation}
describing respectively, $C_3$ rotation, mirror plane, time reversal, and lattice translation.
Here $\Sigma$, $\sigma$,  and $\tau$ denote sub-lattice, spin, and valley Pauli matrices respectively.
Doing so we classify the bilinears of matrices which form invariants under the symmetry group.
We note that $s_x,s_y$ and $\tau_x,\tau_y$ form doublets under $C_3$ and $T$ respectively.
We may now tabulate which matrices are even or odd under $\sigma_V$ and $\theta$ in \cref{tab:sym-int-vert}.
\begin{table}[!htp]
  \centering
  \begin{tabular}{c|c|c}
    &$\theta$ even& $\theta$ odd\\
    \hline
    $\sigma_v$ even& $ \Sigma_y\tau_y$, $ \Sigma_x\tau_z$, $ \Sigma_x\tau_x$, $ \Sigma_z\tau_x$, $ \Sigma_z\tau_z$& $\Sigma_y$, $\tau_y$\\
    $\sigma_v$ odd&$ \Sigma_y\tau_x$, $ \Sigma_y\tau_z$, $ \Sigma_x\tau_y$, $ \Sigma_z\tau_y$&$\Sigma_x$, $\Sigma_z$, $\tau_x$, $\tau_z$
  \end{tabular}
  \caption{Transformation properties of short range interaction vertices, grouped by whether they are even or odd under time reversal $\theta$ and mirror plane $\sigma_V$.\label{tab:sym-int-vert}}
\end{table}
Combining all this we find that the symmetry allowed short ranged interactions are
\begin{multline}
  \mathbbm{1}^1 \cdot \mathbbm{1}^2,
  \bm{\Sigma}^1_\parallel \cdot  \bm{\Sigma}^2_\parallel,
 \bm{\tau}^1_\parallel \cdot \bm{\tau}^2_\parallel,
 \Sigma_z^1 \Sigma^2_z, \tau_z^1 \tau^2_z,\\
  \bm{\Sigma}^1_\parallel \cdot  \bm{\Sigma}^2_\parallel \bm{\tau}^1_\parallel \cdot \bm{\tau}^2_\parallel,
  \bm{\Sigma}^1_\parallel \cdot  \bm{\Sigma}^2_\parallel \tau^1_z \tau^2_z,
 \bm{\tau}^1_\parallel \cdot \bm{\tau}^2_\parallel  \Sigma^1_z  \Sigma^2_z,\\
  \Sigma^1_z \tau^1_z  \Sigma^2_z \tau^2_z,
  \Sigma^1_z \cdot \tau^2_z,
  \Sigma^1_z \tau^1_z \cdot \mathbbm{1}^2
\end{multline}
where we use the superscript $1,2$ to indicates that the vertex is for particle $1,2$, respectively.
We can estimate the interaction constants for these channels by inserting the Dirac cone ansatz for the creation operator
\begin{multline}
 \Psi = u_{KA}\psi_{KA} + u_{KB}\psi_{KB}\\ + u_{K'B}\psi_{K'B} + u_{K'A}\psi_{K'A}
\end{multline}
into the Coulomb interaction.
Because of the relations
\begin{equation}
 u_{KA} = u^*_{K'A},
 u_{KB} = u^*_{K'B},
\end{equation}
the vertices which are odd under time reversal vanish.
The remaining vertices are
\begin{multline}
  \mathbbm{1}^1\cdot\mathbbm{1}^2,
  \bm{\Sigma}^1_\parallel \cdot  \bm{\Sigma}^2_\parallel \bm{\tau}^1_\parallel \cdot \bm{\tau}^2_\parallel,
  \bm{\Sigma}^1_\parallel \cdot  \bm{\Sigma}^2_\parallel \tau^1_z \tau^2_z,\\
 \bm{\tau}^1_\parallel \cdot \bm{\tau}^2_\parallel  \Sigma^1_z  \Sigma^2_z,
  \Sigma^1_z \tau^1_z  \Sigma^2_z \tau^2_z,
  \Sigma^1_z \tau^1_z \cdot \mathbbm{1}^2_z
\end{multline}
to which we assign, respectively, the coupling constants
\begin{equation}
g_{00}, g_{\perp\perp}, g_{z\perp}, g_{\perp z}, g_{zz}, \tilde{g}_{00}.
\end{equation}
Neglecting the overlap of the $A$ and $B$ sub-lattice Bloch wavefunctions we can further discard the $g_{z\perp}, g_{\perp z}$ terms we arrive at 
\begin{multline}
  \hat{H}^\psi_\text{int} = \frac{1}{2} \sum_{\mathbf{r},\mathbf{r}'}
  V(\mathbf{r} - \mathbf{r}') :\psi^\dagger(\mathbf{r}) \psi(\mathbf{r}) \psi^\dagger(\mathbf{r}') \psi(\mathbf{r}'): \\
  + \frac{1}{2} \sum_{\mathbf{r}}\sum_{\alpha\beta}
  \left[g_{\alpha\beta} :\psi^\dagger(\mathbf{r}) \Sigma^\alpha\tau^\beta \psi(\mathbf{r}) \psi^\dagger(\mathbf{r}') \Sigma^\alpha\tau^\beta \psi(\mathbf{r}'):\right.\\
  + \left.
   \tilde{g}_{00} :\psi^\dagger(\mathbf{r}) \Sigma^z\tau^z\psi(\mathbf{r}) \psi^\dagger(\mathbf{r}')\psi(\mathbf{r}'):\right].
\end{multline}

\subsection{Interaction functions in terms of short ranged interaction constants}
Writing the Fermionic annihilation operator
\begin{equation}
  \hat{\Psi}_\sigma(\mathbf{r}) =
  \begin{psmallmatrix}
    u_{KA}(\mathbf{r})&
    u_{KB}(\mathbf{r})&
    u_{K'B}(\mathbf{r})&
    - u_{K'A}(\mathbf{r}) 
  \end{psmallmatrix}
  \cdot 
  \hat{\vec{\psi}}_{\sigma}(\mathbf{r})
\end{equation}
the upper band projected operator has the form $\vec{\psi}_{\mathbf{k}\sigma} = \sum_\zeta \chi_{\mathbf{k}\zeta} c_{\mathbf{k}\zeta\sigma}$
where
\begin{equation}
\begin{gathered}
\chi_{\mathbf{k}+} = 
\begin{pmatrix}
\sqrt{\frac{1}{2}\left(1 + \frac{\Delta}{E_\mathbf{k}}\right)}\\
e^{i\phi_\mathbf{k}}\sqrt{\frac{1}{2}\left(1 - \frac{\Delta}{E_\mathbf{k}}\right)}
\end{pmatrix},\\
\chi_{\mathbf{k}-} = 
-\begin{pmatrix}
e^{-i\phi_\mathbf{k}}\sqrt{\frac{1}{2}\left(1 - \frac{\Delta}{E_\mathbf{k}}\right)}\\
\sqrt{\frac{1}{2}\left(1 + \frac{\Delta}{E_\mathbf{k}}\right)}
\end{pmatrix}.
\end{gathered}
\end{equation}
By plugging the upper band operator into \cref{eq:Hint-sublattice}, the interaction functions in \cref{eq:gint} can then be expressed in terms of the interaction constants $g_{zz},g_{\perp\perp},\widetilde g_{00}$ and the function $V(\mathbf{q})$ combined with matrix elements of the eigenspinors $\chi_{\mathbf{k}\zeta}$.
\begin{widetext}
Using the shorthand
\begin{equation}
F_{ij;i'j'} \equiv F(\mathbf{p}_i, \mathbf{p}_{j}; \mathbf{p}_{i'}, \mathbf{p}_{j'})
\equiv F(\mathbf{p} + \mathbf{q}/2, \mathbf{p}'-\mathbf{q}/2; \mathbf{p}-\mathbf{q}/2, \mathbf{p}' + \mathbf{q}/2)
\end{equation}
we can write
\begin{equation}
  \begin{aligned}
      U^d_{\mathbf{p},\mathbf{p}',\mathbf{q}}&=
      \frac{1}{2}\left(V_{\mathbf{q}}L_{ii'}L_{jj'}
      - \frac{1}{4}V_{\mathbf{p} -\mathbf{p}'}L_{ij'}L_{ji'}
      + g_{zz}\left(N_{ii'}N_{jj'} - \frac{1}{4}N_{ij'}N_{ji'}\right)\right.\\
  &\left.+ \frac{1}{2}\tilde{g}_{00}
      \left(
        Q^{00}_{ii';jj'}
        -\frac{1}{4}Q^{00}_{ij';ji'}
      \right)
      - \frac{1}{2}g_{\perp\perp}Q^{\perp\perp}_{ij';ji'}
      \right)
\\
      U^s_{\mathbf{p},\mathbf{p}',\mathbf{q}} &=
      -\frac{1}{4}\left(
      \frac{1}{2}V_{\mathbf{p} -\mathbf{p}'}L_{ij'}L_{ji'}
      + \frac{1}{2}g_{zz}N_{ij'}N_{ji'}
      + \frac{1}{4}\tilde{g}_{00}Q^{00}_{ij';ji'}
      + g_{\perp\perp}Q^{\perp\perp}_{ij';ji'}
\right)
\\
 U^{v\parallel}_{\mathbf{p},\mathbf{p}',\mathbf{q}} &= 
      -\frac{1}{2} \left(\frac{1}{4}V_{\mathbf{p} -\mathbf{p}'}L_{ij'}L_{ji'}
      + \frac{1}{4}g_{zz}N_{ij'}N_{ji'}
      + \frac{1}{8}\tilde{g}_{00}Q^{00}_{ij';ji'}
      -g_{\perp\perp}Q^{\perp\perp}_{ii';jj'}
      \right)
    \\
      U^{vz/mz}_{\mathbf{p},\mathbf{p}',\mathbf{q}} &=
      -\frac{1}{4}\left(
      \frac{1}{2}V_{\mathbf{p} -\mathbf{p}'}L_{ij'}L_{ji'}
      + \frac{1}{2}g_{zz}N_{ij'}N_{ji'}
      + \frac{1}{4}\tilde{g}_{00}Q^{00}_{ij';ji'}
      - g_{\perp\perp}Q^{\perp\perp}_{ij';ji'}\right)\\
      U^{m\parallel}_{\mathbf{p},\mathbf{p}',\mathbf{q}} &=
      -\frac{1}{8}\left(
      V_{\mathbf{p} -\mathbf{p}'}L_{ij'}L_{ji'}
      + g_{zz}N_{ij'}N_{ji'}
      + \frac{1}{2}\tilde{g}_{00}Q^{00}_{ij';ji'}
        \right)
\end{aligned}
 \label{eq:int-functions-estimate}
\end{equation}
where
\begin{equation}
  \begin{gathered}
 Q^{\perp\perp}_{ii';jj'} =  \tilde{L}_{ii'}\tilde{L}^*_{jj'} + \tilde{N}_{ii'}\tilde{N}^*_{jj'}
, \qquad Q^{00}_{ii';jj'} = L_{ii'}N_{jj'} + N_{ii'}L_{jj'}\\
 L(\mathbf{k},\mathbf{k}') = \chi_{\mathbf{k}+}^\dagger\chi_{\mathbf{k}'+},\qquad N(\mathbf{k},\mathbf{k}') = \chi_{\mathbf{k}+}^\dagger\tau_3\chi_{\mathbf{k}'+},\qquad
 \tilde{L}(\mathbf{k},\mathbf{k}') = \chi_{\mathbf{k}+}^\dagger\chi^*_{\mathbf{k}'+},\qquad
 \tilde{N}(\mathbf{k},\mathbf{k}') = \chi_{\mathbf{k}+}^\dagger\tau_3\chi^*_{\mathbf{k'}+}
\end{gathered}
\label{eq:coherence}
\end{equation}
\end{widetext}
Of particular note are the limits of the coherence factors $L, N, \tilde{L}, \tilde{N}$,
\begin{equation}
\begin{gathered}
L(\mathbf{k}, \mathbf{k}) = N(\mathbf{k}, - \mathbf{k}) = 1\\
L(\mathbf{k}, -\mathbf{k}) = N(\mathbf{k},  \mathbf{k}) = \frac{\Delta}{E_{\mathbf{k}}}, 
\end{gathered}
\end{equation}
where the latter relation is responsible for the suppression of backscattering in the relativistic, $\Delta \to 0$ limit.

\section{Low energy interactions from the Coulomb interaction}
\label{sec:int-coulomb}
In this section, we estimate the strength of short ranged interactions in the various spin-valley channels.

We can obtain estimates for the interactions strengths occurring in \cref{eq:Hint-sublattice,eq:int-functions-estimate} by looking at the matrix elements of the Coulomb interaction.
In particular we aim to obtain the lowest harmonic of the Coulomb interaction contributing to each interaction strength above.

We begin by writing the interaction Hamiltonian for the physical electrons in the bands nearest the Fermi surface.
\begin{multline}
  \hat{H}_\text{int} = \frac{1}{2} \sum_{\lambda\lambda',\sigma\sigma'} \int d\mathbf{r}  d\mathbf{r}' V(\abs{\mathbf{r} - \mathbf{r}'})\\
  \times
  :c^\dagger_{\sigma\lambda}(\mathbf{r})c_{\sigma\lambda}(\mathbf{r}) c^\dagger_{\sigma'\lambda'}(\mathbf{r}')c_{\sigma'\lambda'}(\mathbf{r}') :
  \label{eq:Hcoulomb}
\end{multline}
where $\sigma$ denotes spin, $\lambda=\pm1$ denotes the sublattice A and B states respectively, $:\cdots:$ denotes normal ordering, and $V(\abs{\mathbf{R}})$ is the Coulomb interaction.
The operators may be expanded in terms of Bloch states
\begin{equation}
 c_{\sigma\lambda}(\mathbf{r}) = \int_\text{BZ}\frac{d\mathbf{p}}{{(2\pi)}^2} \psi_{\lambda\mathbf{p}}(\mathbf{r}) c_{\sigma\lambda}(\mathbf{p})
\end{equation}
with wavefunction
\begin{equation}
 \psi_{\lambda\mathbf{p}}(\mathbf{r}) = e^{i\mathbf{p} \cdot\mathbf{r}}u_{\lambda\mathbf{p}}(\mathbf{r}).
\end{equation}
Using the Bloch basis and the Fourier transform of the Coulomb interaction we can rewrite \cref{eq:Hcoulomb}
\begin{multline}
  \hat{H}_\text{int} = \frac{1}{2} \sum_{\lambda\lambda',\sigma\sigma'}
  \int_{\text{BZ}}\left[\prod_i \frac{d\mathbf{p}_i}{{(2\pi)}^2}\right]
  \int_{\mathbb{R}^2}\frac{d\mathbf{q}}{{(2\pi)}^2}\\
  \times
  V(q)
  :c^\dagger_{\sigma\lambda}(\mathbf{p}_1)c_{\sigma\lambda}(\mathbf{p}_1') c^\dagger_{\sigma'\lambda'}(\mathbf{p}_2)c_{\sigma'\lambda'}(\mathbf{p}_2') :\\
  \times\int_{\mathbb{R}^2} d\mathbf{r}  d\mathbf{r}' e^{i\mathbf{q}\cdot (\mathbf{r} - \mathbf{r}')}\\
  \times
  \psi^*_{\lambda\mathbf{p}_1}(\mathbf{r})  \psi_{\lambda\mathbf{p}_1'}(\mathbf{r})
  \psi^*_{\lambda\mathbf{p}_2}(\mathbf{r}')  \psi_{\lambda\mathbf{p}_2'}(\mathbf{r}').
\end{multline}
We can rewrite the integral of $\mathbf{q}$ as integral over the Brillouin zone and a sum over all reciprocal lattice vactors
\begin{equation}
 \int_{\mathbb{R}^2}  \frac{d\mathbf{q}}{{(2\pi)}^2} f(\mathbf{q}) = \sum_{\mathbf{G}}  \int_{BZ}  \frac{d\mathbf{q}}{{(2\pi)}^2} f(\mathbf{q} + \mathbf{G})
\end{equation}
where we have introduced the short-hand
\begin{equation}
 \sum_\mathbf{G} f(\mathbf{G}) \equiv \sum_{n_1,n_2} f(n_1 \mathbf{b}_1 + n_2 \mathbf{b}_2)
\end{equation}
with $\mathbf{b}_i$ the primitive reciprocal lattice vectors.
Similarly we can rewrite the integration over all space in terms of integration over the unit cell
\begin{equation}
 \int_{\mathbf{R}^2} d\mathbf{r} f(\mathbf{r}) = \sum_\mathbf{R} \int_\text{uc} d\mathbf{r} f(\mathbf{r} + \mathbf{R})
\end{equation}
where we have analogously defined the shorthand
\begin{equation}
 \sum_\mathbf{R} f(\mathbf{R}) \equiv \sum_{n_1,n_2} f(n_1 \mathbf{a}_1 + n_2 \mathbf{a}_2)
\end{equation}
with $\mathbf{a}_i$ the primitive translation vectors of the lattice.
\begin{widetext}
Using both these relations, the interaction hamiltonian becomes
\begin{multline}
  \hat{H}_\text{int} = \frac{1}{2} \sum_{\lambda\lambda',\sigma\sigma'}
  \sum_{\mathbf{G}} \int_{\text{BZ}}\left[\prod_i \frac{\mathbf{p}_i}{{(2\pi)}^2}\right]
  \int_{\text{BZ}}\frac{d\mathbf{q}}{{(2\pi)}^2}
  :c^\dagger_{\sigma\lambda}(\mathbf{p}_1)c_{\sigma\lambda}(\mathbf{p}_1') c^\dagger_{\sigma'\lambda'}(\mathbf{p}_2)c_{\sigma'\lambda'}(\mathbf{p}_2') :\\
  \times
    V(\abs{\mathbf{q} + \mathbf{G}})
    \sum_{\mathbf{R},\mathbf{R}'} e^{i\mathbf{q} \cdot (\mathbf{R} - \mathbf{R}')}
    e^{-i(\mathbf{p}_1 - \mathbf{p}_1')\cdot \mathbf{R}}
    e^{-i(\mathbf{p}_2 - \mathbf{p}_2')\cdot \mathbf{R}'}\\
    \times
    \int_{\text{uc}} d\mathbf{r}  d\mathbf{r}' e^{i\left(\mathbf{q} + \mathbf{G}\right)\cdot (\mathbf{r} - \mathbf{r}')}
  \psi^*_{\lambda\mathbf{p}_1}(\mathbf{r})  \psi_{\lambda\mathbf{p}_1'}(\mathbf{r})
  \psi^*_{\lambda\mathbf{p}_2}(\mathbf{r}')  \psi_{\lambda\mathbf{p}_2'}(\mathbf{r}').
\end{multline}
where we have used $\exp(i \mathbf{G} \cdot \mathbf{R}) = 1$, and the periodicity properties of the Bloch wavefunctions.
The sums over $\mathbf{R}, \mathbf{R}'$ can be performed to obtained
\begin{equation}
{(2\pi)}^2\delta_{BZ}(\mathbf{p}_1 - \mathbf{p}_1' - \mathbf{q})
{(2\pi)}^2\delta_{BZ}(\mathbf{p}_2 - \mathbf{p}_2' + \mathbf{q})
\end{equation}
where $\delta_{BZ}$ is to be resolved modulo the Brilloun zone, as the integrand is periodic in all $\mathbf{p}_i$.
We then obtain
\begin{multline}
  \hat{H}_\text{int} = \frac{1}{2} \sum_{\lambda\lambda',\sigma\sigma'}
  \sum_{\mathbf{G}} \int_{\text{BZ}}\frac{d\mathbf{p}}{{(2\pi)}^2}\frac{d\mathbf{p}'}{{(2\pi)}^2}
  \int_{\text{BZ}}\frac{d\mathbf{q}}{{(2\pi)}^2}
  :c^\dagger_{\sigma\lambda}(\mathbf{p}+\mathbf{q})c_{\sigma\lambda}(\mathbf{p}) c^\dagger_{\sigma'\lambda'}(\mathbf{p}'-\mathbf{q})c_{\sigma'\lambda'}(\mathbf{p}') :\\
  \times
    V(\abs{\mathbf{q} + \mathbf{G}})
    \int_{\text{uc}} d\mathbf{r}  d\mathbf{r}' e^{i\left(\mathbf{q} + \mathbf{G}\right)\cdot (\mathbf{r} - \mathbf{r}')}
  \psi^*_{\lambda\mathbf{p}+\mathbf{q}}(\mathbf{r})  \psi_{\lambda\mathbf{p}}(\mathbf{r})
  \psi^*_{\lambda\mathbf{p}'-\mathbf{q}}(\mathbf{r}')  \psi_{\lambda\mathbf{p}'}(\mathbf{r}')
\end{multline}
and in terms of the Bloch factors
\begin{multline}
  \hat{H}_\text{int} = \frac{1}{2} \sum_{\lambda\lambda',\sigma\sigma'}
  \sum_{\mathbf{G}} \int_{\text{BZ}}\frac{d\mathbf{p}}{{(2\pi)}^2}\frac{d\mathbf{p}'}{{(2\pi)}^2}
  \int_{\text{BZ}}\frac{d\mathbf{q}}{{(2\pi)}^2}
  :c^\dagger_{\sigma\lambda}(\mathbf{p}+\mathbf{q})c_{\sigma\lambda}(\mathbf{p}) c^\dagger_{\sigma'\lambda'}(\mathbf{p}'-\mathbf{q})c_{\sigma'\lambda'}(\mathbf{p}') :\\
  \times
    V(\abs{\mathbf{q} + \mathbf{G}})
    \int_{\text{uc}} d\mathbf{r}  d\mathbf{r}' e^{i\mathbf{G}\cdot (\mathbf{r} - \mathbf{r}')}
  u^*_{\lambda\mathbf{p}+\mathbf{q}}(\mathbf{r})  u_{\lambda\mathbf{p}}(\mathbf{r})
  u^*_{\lambda\mathbf{p}'-\mathbf{q}}(\mathbf{r}')  u_{\lambda\mathbf{p}'}(\mathbf{r}').
\end{multline}
We may thus write the interaction as
\begin{equation}
  \hat{H}_\text{int} = \frac{1}{2} \sum_{\lambda\lambda',\sigma\sigma'}
  \int_{\text{BZ}}\frac{d\mathbf{p}}{{(2\pi)}^2}\frac{d\mathbf{p}'}{{(2\pi)}^2}
  \int_{\text{BZ}}\frac{d\mathbf{q}}{{(2\pi)}^2}
  U_{\lambda\lambda'}(\mathbf{p}, \mathbf{p}', \mathbf{q})
  :c^\dagger_{\sigma\lambda}(\mathbf{p}+\mathbf{q})c_{\sigma\lambda}(\mathbf{p}) c^\dagger_{\sigma'\lambda'}(\mathbf{p}'-\mathbf{q})c_{\sigma'\lambda'}(\mathbf{p}') :
\end{equation}
where we have defined
\begin{equation}
  U_{\lambda\lambda'}(\mathbf{p}, \mathbf{p}', \mathbf{q}) \equiv
    \sum_{\mathbf{G}} V(\abs{\mathbf{q} + \mathbf{G}})
    \int_{\text{uc}} d\mathbf{r}  d\mathbf{r}' e^{i\mathbf{G}\cdot (\mathbf{r} - \mathbf{r}')}
  u^*_{\lambda\mathbf{p}+\mathbf{q}}(\mathbf{r})  u_{\lambda\mathbf{p}}(\mathbf{r})
  u^*_{\lambda\mathbf{p}'-\mathbf{q}}(\mathbf{r}')  u_{\lambda\mathbf{p}'}(\mathbf{r}').
\end{equation}

Now we make the Dirac cone approximation.
We restrict the momenta of each of the electron operators to be in the vicinity of the $\mathbf{K}$ or $\mathbf{K}'$ point.
Each of the Bloch factors will be evaluated at the corresponding point.
Introducing the notation
\begin{equation}
 \int_\Lambda \frac{d\mathbf{p}}{{(2\pi)}^2} \dotsi
\end{equation}
for integration over momementa $\mathbf{p} \ll \abs{\mathbf{K} - \mathbf{K}'}$ and the operators
\begin{equation}
 c_{\sigma\lambda\zeta}(\mathbf{p}) = c_{\sigma \lambda}(\zeta \mathbf{K}  + \mathbf{p})
\end{equation}
with $\zeta= \pm1$ indexing the valley degree of freedom, we may approximate the interaction Hamiltonian as
\begin{equation}
  \hat{H}_\text{int} \approx \frac{1}{2} \sum_{\lambda\lambda',\sigma\sigma',\zeta_i}
  \int_{\Lambda}\frac{d\mathbf{p}}{{(2\pi)}^2}\frac{d\mathbf{p}'}{{(2\pi)}^2}
  \int_{\Lambda} \frac{d\mathbf{q}}{{(2\pi)}^2}
  U_{\lambda\lambda';\zeta_i}(\mathbf{q})
  :c^\dagger_{\sigma\lambda\zeta_1}(\mathbf{p}+\mathbf{q})c_{\sigma\lambda\zeta_1'}(\mathbf{p}) c^\dagger_{\sigma'\lambda'\zeta_2}(\mathbf{p}'-\mathbf{q})c_{\sigma'\lambda'\zeta_2'}(\mathbf{p}') :
\end{equation}
with
\begin{equation}
  U_{\lambda\lambda';\zeta_i}(\mathbf{q}) \equiv
    \sum_{\mathbf{G}} V(\abs{\mathbf{q} + \mathbf{G} + (\zeta_1 - \zeta_1')\mathbf{K}})
    \delta_{\zeta_1 - \zeta_1',\zeta_2' - \zeta_2}
    \int_{\text{uc}} d\mathbf{r}  d\mathbf{r}' e^{i\mathbf{G}\cdot (\mathbf{r} - \mathbf{r}')}
  u^*_{\lambda\zeta_1}(\mathbf{r})  u_{\lambda\zeta_1'}(\mathbf{r})
  u^*_{\lambda\zeta_2}(\mathbf{r}')  u_{\lambda\zeta_2'}(\mathbf{r}').
\end{equation}
There are two types of terms, $\zeta_1 = \zeta_1' = \zeta, \zeta_2 = \zeta_2' = \zeta'$ and $\zeta_1 = - \zeta_1' = \zeta_2' = -\zeta_2 = \zeta$.
We therefore write
\begin{multline}
  \hat{H}_\text{int} \approx
  \frac{1}{2} \sum_{\lambda\lambda',\sigma\sigma', \zeta\zeta'}
  \int_{\Lambda}\frac{d\mathbf{p}}{{(2\pi)}^2}\frac{d\mathbf{p}'}{{(2\pi)}^2}
  \int_{\Lambda} \frac{d\mathbf{q}}{{(2\pi)}^2}
  U^\text{intra}_{\lambda\lambda'}(\mathbf{q})
  :c^\dagger_{\sigma\lambda\zeta}(\mathbf{p}+\mathbf{q})c_{\sigma\lambda\zeta}(\mathbf{p}) c^\dagger_{\sigma'\lambda'\zeta'}(\mathbf{p}'-\mathbf{q})c_{\sigma'\lambda'\zeta'}(\mathbf{p}') :\\
  +
  \frac{1}{2} \sum_{\lambda\lambda',\sigma\sigma', \zeta}
  \int_{\Lambda}\frac{d\mathbf{p}}{{(2\pi)}^2}\frac{d\mathbf{p}'}{{(2\pi)}^2}
  \int_{\Lambda} \frac{d\mathbf{q}}{{(2\pi)}^2}
  U^\text{inter}_{\lambda\lambda';\zeta}(\mathbf{q})
  :c^\dagger_{\sigma\lambda\zeta}(\mathbf{p}+\mathbf{q})c_{\sigma\lambda,-\zeta}(\mathbf{p}) c^\dagger_{\sigma'\lambda',-\zeta}(\mathbf{p}'-\mathbf{q})c_{\sigma'\lambda'\zeta}(\mathbf{p}') :.
\end{multline}
with
\begin{equation}
  U^\text{intra}_{\lambda\lambda'}(\mathbf{q}) \equiv
    \sum_{\mathbf{G}} V(\abs{\mathbf{q} + \mathbf{G}})
    \int_{\text{uc}} d\mathbf{r}  d\mathbf{r}' e^{i\mathbf{G}\cdot (\mathbf{r} - \mathbf{r}')}
  \abs{u_{\lambda+}(\mathbf{r})}^2
  \abs{u_{\lambda+}(\mathbf{r}')}^2
\end{equation}
where we have used that fact that $u_{-} = u^*_+ \implies \abs{u_+}^2 = \abs{u_-}^2$
and
\begin{equation}
  U^\text{inter}_{\lambda\lambda';\zeta}(\mathbf{q}) \equiv
    \sum_{\mathbf{G}} V(\abs{\mathbf{q} + \mathbf{G} + 2 \zeta\mathbf{K}})
    \int_{\text{uc}} d\mathbf{r}  d\mathbf{r}' e^{i\mathbf{G}\cdot (\mathbf{r} - \mathbf{r}')}
  u^*_{\lambda\zeta}(\mathbf{r})  u_{\lambda,-\zeta}(\mathbf{r})
  u^*_{\lambda,-\zeta}(\mathbf{r}')  u_{\lambda,\zeta}(\mathbf{r}').
\end{equation}
We then see that $U^\text{intra}$ corresponds to $g_{00}, \tilde{g}_{00}, g_{zz}$ and $U^\text{inter}$ to $g_{\perp\perp}$.
From this we can immediately deduce that the smallest harmonic of the potential contributing to $g_{\perp\perp}$ is $V(|\mathbf{K} - \mathbf{K}'|)$.
For $U^\text{intra}$ we rewrite
\begin{equation}
 \abs{u_\lambda}^2 = \frac{1}{2}\left(\abs{u_A}^2 + \abs{u_B}^2\right) + \frac{\lambda}{2}\left(\abs{u_A}^2 - \abs{u_B}^2\right)
\end{equation}
giving
\begin{multline}
  U^\text{intra}_{\lambda\lambda'}(\mathbf{q}) \equiv
    \frac{1}{4} \sum_{\mathbf{G}} V(\abs{\mathbf{q} + \mathbf{G}})
    \left\{
   \abs{ \int_{\text{uc}} d\mathbf{r} {r}' e^{i\mathbf{G}\cdot \mathbf{r} }
        \left(\abs{u_{A+}(\mathbf{r})}^2 + \abs{u_{B+}(\mathbf{r})}^2\right)}^2\right.\\
   + \lambda\lambda'\abs{ \int_{\text{uc}} d\mathbf{r}  e^{i\mathbf{G}\cdot \mathbf{r} }
        \left(\abs{u_{A+}(\mathbf{r})}^2 - \abs{u_{B+}(\mathbf{r})}^2\right)}^2\\
        + \lambda
    \int_{\text{uc}} d\mathbf{r}  d\mathbf{r}' e^{i\mathbf{G}\cdot (\mathbf{r} - \mathbf{r}')}
\left(\abs{u_{A+}(\mathbf{r})}^2 - \abs{u_{B+}(\mathbf{r})}^2\right)
\left(\abs{u_{A+}(\mathbf{r}')}^2 + \abs{u_{B+}(\mathbf{r}')}^2\right)\\
      \left.
        + \lambda'
    \int_{\text{uc}} d\mathbf{r}  d\mathbf{r}' e^{i\mathbf{G}\cdot (\mathbf{r} - \mathbf{r}')}
\left(\abs{u_{A+}(\mathbf{r})}^2 + \abs{u_{B+}(\mathbf{r})}^2\right)
\left(\abs{u_{A+}(\mathbf{r}')}^2 - \abs{u_{B+}(\mathbf{r}')}^2\right)
      \right\}.
\end{multline}
The first and second terms can be identified with $g_{00}$ and $g_{zz}$ respectively, while he third and fourth are $\tilde{g}_{00}$
It is clear that in the presence of sublattice symmetry the third and fourth terms must vanish as they are odd under the exchange of $A$ and $B$.
Using the fact that the Bloch factors are normalized to $1$ we see that the lowest order contribution to $g_{00}$ is
\begin{equation}
 g_{00} = V(q) + \text{higher harmonics}
\end{equation}
while the $\mathbf{G}=0$ contributions for the other terms all vanish since
\begin{equation}
 \int_\text{uc} d\mathbf{r}
\left(\abs{u_{A+}(\mathbf{r})}^2 - \abs{u_{B+}(\mathbf{r})}^2\right) = 1-1 =0.
\end{equation}
The lowest contributing harmonics are
\begin{equation}
  \begin{gathered}
    g_{00} \approx V(q)\\
    \begin{multlined}[][\arraycolsep]
     g_{\perp\perp}\approx
    \frac{1}{2}V\left(\frac{4\pi}{3a}\right) \sum_{i}
    \left(
      {\left|\int_{uc}d\mathbf{r}
    \cos(\mathbf{b}_i \cdot \mathbf{r})
  \left({\left[u_{A+}(\mathbf{r})\right]}^2
  + {\left[u_{B+}(\mathbf{r})\right]}^2\right)
    \right|}^2\right.\\
\left.
+ {\left|\int_{uc}d\mathbf{r}
    \sin(\mathbf{b}_i \cdot \mathbf{r})
\left({\left[u_{A+}(\mathbf{r})\right]}^2
  + {\left[u_{B+}(\mathbf{r})\right]}^2\right)
\right|}^2\right)
    \end{multlined}
    \\
    \begin{multlined}[][\arraycolsep]
    g_{zz} \approx \frac{1}{2}V\left(\frac{4\pi}{\sqrt{3}a}\right)\sum_{i=1,2,3}\left(
      {\left[\int_{uc}d\mathbf{r}
    \cos(\mathbf{b}_i \cdot \mathbf{r})
    \left(\abs{u_{A+}(\mathbf{r})}^2 - \abs{u_{B+}(\mathbf{r})}^2\right)\right]}^2\right.\\
\left.
+ {\left[\int_{uc}d\mathbf{r}
    \sin(\mathbf{b}_i \cdot \mathbf{r})
\left(\abs{u_{A+}(\mathbf{r})}^2 - \abs{u_{B+}(\mathbf{r})}^2\right)\right]}^2\right)
\end{multlined}\\
    \tilde{g}_{00} \approx
    \sum_{i=1,2,3} V\left(\frac{4\pi}{\sqrt{3}a}\right)
    \int_{\text{uc}} d\mathbf{r}  d\mathbf{r}' \cos(\mathbf{b}_i\cdot (\mathbf{r} - \mathbf{r}'))
\left(\abs{u_{A+}(\mathbf{r})}^2\abs{u_{A+}(\mathbf{r}')}^2  - \abs{u_{B+}(\mathbf{r})}^2\abs{u_{B+}(\mathbf{r}')}^2\right)
  \end{gathered}
  \label{eq:interaction-lowest}
\end{equation}
\end{widetext}
where $\mathbf{b}_{1,2}$ are the primitive reciprocal lattice vectors and $\mathbf{b}_3 = \mathbf{b}_1 + \mathbf{b}_2$.
In what follows we therefore, set $g_{\perp\perp},g_{zz},\tilde{g}_{00}$ to constants and replace $g_{00}$ with the long range part of the Coulomb interaction.
Note that as a consequence of the monotonicity of the Coulomb potential,
for the hierarchy of scales $q_{TF}, k_F \ll \abs{\mathbf{K} - \mathbf{K}'}$ we then have
\begin{equation}
 V(q \sim k_F) \gg g_{\perp\perp} > g_{zz} > \tilde{g}_{00}.
\end{equation}

\section{Relaxation rates for arbitrary angular harmonic}
\label{sec:relax-arb}
With illustrate here the evaluation of \cref{eq:Imuminus}.  At low temperatures, the particles are restricted to the Fermi surface and we can write the scattering rates as $W^\mu_{-}(\phi - \phi', \theta)$, 
where we have defined $\phi,\phi'$ as the angles of $\mathbf{p}_{i} + \mathbf{p}_{i'}$ and $\mathbf{p}_j + \mathbf{p}_{j'}$ respectively and
where $\theta = \sin^{-1}(q/2k_F)$ is the scattering angle.
We also reparametrize the linearized deviations in terms of angular harmonics on the Fermi surface
\begin{equation}
   \delta \bar{\rho}^\mu(\mathbf{p}_i, \mathbf{r}) 
   \equiv -\left.\frac{\partial n}{\partial\epsilon}\right|_{\bar{\epsilon}}\sum_m e^{im\phi_i}\eta^\mu_m(\mathbf{r}).
\end{equation}
Plugging this form into \cref{eq:Imuminus} allows us to straightforwardly obtain the angular harmonics of the collision integral
\begin{widetext}
\begin{multline}
  I^\mu_m[\eta^\mu] \equiv \nu_F^{-1}\sum_\mathbf{p} e^{-im \phi} I^\mu_i[\mathbf{p}]
  = -\frac{1}{\nu_F T} \int \frac{d^2p_i d^2p_j d^2p_{i'} d^2p_{j'}}{{(2\pi)}^5}
  \delta(\sum'_J \mathbf{p}_{J})\delta(\sum'_J \epsilon_J)
n_i n_j(1-n_{i'})(1-n_{j'})\\
\times
   \sum_{\pm, m'} e^{-im\phi_i}
   W^\mu_{\pm}(\phi - \phi', \theta)\left(
   \eta^\mu_{i,m'} e^{im'\phi_i} \pm \eta^\mu_{j,m'}e^{im'\phi_j} -  \eta^\mu_{i', m'} e^{im'\phi_{i'}} \mp \eta^\mu_{j', m'} e^{im'\phi_{j'}}\right)
 \label{eq:Iharmonics}
\end{multline}
The probabilities $W_\pm$ entering the collision integral are defined respectively as the probabilities due to scattering of particles with the same quantum number in the associated channel,
which are important only for the even modes and the charge channel, and the scattering of particles with different quantum number in the associated channel.
For example, for the spin mode, $W_+$ describes scattering of particles with the same spin, while $W_-$ describes scattering of particles with opposite spin.

Explicitly,
\begin{equation}
  \begin{gathered}
  W^d_+ =W_{\uparrow\uparrow;++} + 2W^D_{\uparrow\uparrow;+-}
  + 2W^D_{\uparrow\downarrow;++}
  + 2W^D_{\uparrow\downarrow;+-},\quad
 W^s_+ = W_{\uparrow\uparrow;++} + 2W^D_{\uparrow\uparrow;+-}, \quad
 W^{vz}_+ = W_{\uparrow\uparrow;++} + 2W^D_{\uparrow\downarrow;++}\\
 W^{mz}_+ =W_{\uparrow\uparrow;++} + 2W^D_{\uparrow\downarrow;+-},\quad
 W^{v\parallel}_+ =W_{\uparrow\uparrow;++} + 2W^{xD}_{\uparrow\downarrow;++},\quad
 W^{m\parallel}_+ = W_{\uparrow\uparrow;++} + 2W^{xD}_{\uparrow\downarrow;+-}
\end{gathered}
\end{equation}
and 
\begin{equation}
  \begin{gathered}
 W^s_- = 2(W^D_{\uparrow\downarrow;++}  + W^D_{\uparrow\downarrow;+-}),\quad
 W^{vz}_- = 2(W^D_{\uparrow\uparrow;+-} + W^D_{\uparrow\downarrow;+-}),\quad
 W^{mz}_- = 2(W^D_{\uparrow\uparrow;+-} + W^D_{\uparrow\downarrow;++})\\
 W^{v\parallel}_- = 2(W^{xD}_{\uparrow\uparrow;+-} + W^{xD}_{\uparrow\downarrow;+-}),\quad
 W^{m\parallel}_- = 2(W^{xD}_{\uparrow\uparrow;+-} + W^{xD}_{\uparrow\downarrow;++}).
\end{gathered}
\end{equation}
where $W^D_{\sigma\sigma';\zeta\zeta'}$ is the scattering probability for two \emph{distinguishable} particles~\cite{Emery1968,Nozieres1999} with spins $\sigma,\sigma'$ and valley indices $\zeta,\zeta'$, $W_{\uparrow\uparrow;++}$ is the scattering probability for two \emph{indistinguishable} particles in the same spin and valley states, and the superscript $x$ indicates the choice of the $\tau^x$ eigenbasis in valley space.

We follow here the methodology of Ref.~\onlinecite{Ledwith2019}, wherein the integral is evaluated for the $+$ terms (consequently we do not repeat the derivation of the $+$ terms here).
First we rewrite the delta functions
\begin{equation}
 \delta(\epsilon_i + \epsilon_{j} - \epsilon_{i'} - \epsilon_{j'})= \int d\omega \delta(\epsilon_i - \epsilon_{i'} - \omega) \delta(\epsilon_{j}  - \epsilon_{j'} + \omega)
 \label{eq:deltaomega}
\end{equation}
and
\begin{equation}
 \delta(\mathbf{p}_i + \mathbf{p}_{j} - \mathbf{p}_{i'} - \mathbf{p}_{j'})
 = \int d^2\!q \delta(\mathbf{p}_i - \mathbf{p}_{i'} - \mathbf{q}) \delta(\mathbf{p}_{j}  - \mathbf{p}_{j'} + \mathbf{q}).
\end{equation}
Changing variables to
\begin{equation}
 \mathbf{p}_i,\mathbf{p}_{i'} = \mathbf{p}  + \frac{\mathbf{l}}{2},
 \qquad \mathbf{p}_j,\mathbf{p}_{j'} = \mathbf{p}'  - \frac{\mathbf{l}'}{2},
\end{equation}
we can resolve the delta functions to find $\mathbf{l} = \mathbf{l}' = \mathbf{q}$.
We rewrite the momentum integrals in polar coordinates
\begin{equation}
 \int d^2{p} = \int dp p\oint d\phi
\end{equation}
This transforms the collision integral to
\begin{multline}
  I^\mu_{m,-}[\eta^\mu]
  = -\frac{1}{\nu_F T} \frac{1}{{(2\pi)}^5}
  \sum_{m}\eta^\mu_{m'}
  \int dp p dp' p' d\omega
  \int dq q
  \oint d\phi d\phi' d\phi_q
  \delta(\epsilon_i - \epsilon_{i'} - \omega)\delta(\epsilon_j - \epsilon_{j'} + \omega)\\
  \times
n_i n_j(1-n_{i'})(1-n_{j'})
    e^{-im\phi_i}W^\mu_{-}(\phi -\phi', \theta)
   \left(
    e^{im'\phi_i}-   e^{im'\phi_{i'}} -e^{im'\phi_j} +   e^{im'\phi_{j'}}\right).
\end{multline}
The difference of energies appearing in the delta functions can be written
\begin{multline}
  \epsilon_i - \epsilon_{i'} =
  \epsilon(\mathbf{p} + \mathbf{q}/2) - \epsilon(\mathbf{p} - \mathbf{q}/2)\\
  =
    \sqrt{v^2p^2 + \frac{v^2q^2}{4} + v^2qp\cos(\phi - \phi_q) + \Delta^2}
    - \sqrt{v^2p^2 + \frac{v^2q^2}{4} - v^2qp\cos(\phi - \phi_q) + \Delta^2}
    \\
    =  \epsilon \left[
    \sqrt{1 + \frac{v^2qp}{\epsilon^2}\cos(\phi - \phi_q)}
    - \sqrt{1 - \frac{v^2qp}{\epsilon^2}\cos(\phi - \phi_q)}
      \right] \equiv \delta\epsilon,
      \label{eq:epsdiff}
\end{multline}
where we have defined
\begin{equation}
  \epsilon \equiv \sqrt{v^2 p^2 + \Delta^2 + \frac{v^2q^2}{4}}.
  \label{eq:epsdef}
\end{equation}
The delta function may then be written
 $\delta(\delta\epsilon - \omega)$ .
Projected to the Fermi surface, the convexity of the square-root implies that
\begin{equation}
 \delta\epsilon(\phi) = 0 \implies \cos(\phi - \phi_q) = 0.
\end{equation}
This allows us to approximate the delta-functions
\begin{equation}
\begin{gathered}
 \delta(\delta\epsilon- \omega)  \approx \frac{1}{\abs{\partial\delta\epsilon/\partial\cos(\phi-\phi_q)}} \delta(\cos(\phi - \phi_q)) \approx \sum_{\chi=\pm} \frac{\mu}{v^2 q p \abs{\sin(\phi - \phi_q)}}\delta(\phi - \phi_q -\chi\frac{\pi}{2})\\
 \delta(\delta\epsilon'- \omega)  \approx \frac{1}{\abs{\partial\delta\epsilon'/\partial\cos(\phi'-\phi_q)}} \delta(\cos(\phi' - \phi_q)) \approx \sum_{\chi'=\pm} \frac{\mu}{v^2 q p \abs{\sin(\phi' - \phi_q)}}\delta(\phi' - \phi_q -\chi\frac{\pi}{2})
\end{gathered}\label{eq:FSdelta}
\end{equation}
The Fermi surface value of $p$ can be obtained from
\begin{equation}
 p^2 = \frac{\mu^2 - \Delta^2 - v^2\frac{q^2}{4}}{v^2}
 = p_F^2 \left(1 - \frac{q^2}{4p_F^2}\right).
\end{equation}
Note that we use $\mu$ for chemical potential for compactness of notation here, which is related to Fermi level in the main text by $\mu = E_F + \Delta$.
This along with, $v_F = \frac{v^2 p_F}{\mu}$
allows us to simplify \cref{eq:FSdelta} to
\begin{equation}
\begin{gathered}
 \delta(\delta\epsilon- \omega)   \approx \sum_{\chi=\pm} \frac{1}{v_F q \sqrt{1- {(q/2p_F)}^2}}\delta(\phi - \phi_q -\chi\frac{\pi}{2})\\
 \delta(\delta\epsilon'- \omega)   \approx \sum_{\chi'=\pm} \frac{1}{v_F q \sqrt{1- {(q/2p_F)}^2}}\delta(\phi' - \phi_q -\chi'\frac{\pi}{2}).
\end{gathered}\label{eq:FSsimp}
\end{equation}

Putting this back in
\begin{multline}
  I^\mu_{m,-}[\eta^\mu]
  = -\frac{1}{\nu_F v_F^2 T} \frac{1}{{(2\pi)}^5}
  \sum_{m'}\eta^\mu_{m'}
  \int dp p dp p' d\omega
  \int \frac{dq}{q (1 - {(q/2p_F)}^2)}\\
  \times
  \sum_{\chi,\chi'}\oint d\phi_q
n_i n_j(1-n_{i'})(1-n_{j'})
  e^{-im\phi_i}
W^\mu_{-}((\chi - \chi')\frac{\pi}{2}, \theta)
   \left(
    e^{im'\phi_i}-   e^{im'\phi_{i'}} -e^{im'\phi_j} +   e^{im'\phi_{j'}}\right).
\end{multline}
\end{widetext}
We now change integration variables to
\begin{equation}
  \epsilon = \sqrt{v^2p^2 + \frac{q^2}{4} + \Delta^2},\,
 \epsilon' = \sqrt{v^2p'^2 + \frac{q^2}{4} + \Delta^2}
\end{equation}
rescale the energy variables by the temperature
\begin{equation}
 \epsilon = \mu + u T,\qquad \epsilon' = \mu + u'T,\qquad \omega = w T
\end{equation}
and change variables for $q$ to
\begin{multline}
  \sin\theta = \frac{q}{2k_F} \implies \frac{dq}{q(1 - {(q/2p_F)}^2)} = \frac{d\theta \abs{\cos\theta}}{\sin\theta\cos^2\theta}\\
  = \frac{d\theta}{\sin\theta\cos\theta}.
\end{multline}
Then to leading order this renders the collision integral
\begin{multline}
  I^\mu_{m,-}[\eta^\mu]
  = -
  \sum_{m'}\eta^\mu_{m'}
  \frac{\nu_F T^2}{v^2_F}\int d\Sigma\\
  \times
  \int_0^{\pi/2} \frac{d\theta}{\sin\theta\cos\theta}
  \sum_{\chi,\chi'}\oint \frac{d\phi_q}{2\pi}
    e^{-im\phi_i}
    W^\mu_{-}((\chi - \chi')\frac{\pi}{2}, \theta)\\
    \times
   \left(
    e^{im'\phi_i}-   e^{im'\phi_{i'}} -e^{im'\phi_j} +   e^{im'\phi_{j'}}\right).
\end{multline}
where for compactness we have defined the integration measure
\begin{equation}
 d\Sigma =
  \frac{1}{4\pi^2} 
  du du' dw
n_i n_j (1 - n_{i'}) (1-n_{j'})
\end{equation}

We now turn to re-expressing the angles $\phi_\alpha$ in terms of $\phi_q$ and $\theta$.
In general, we have
\begin{gather}
 \mathbf{p}_i - \mathbf{p}_{i'} = \mathbf{q}  = \mathbf{p}_{j'} - \mathbf{p}_{j}
\end{gather}
So
\begin{multline}
\sin\abs{\frac{\phi_i - \phi_{i'}}{2}} = \frac{q}{2k_F} = \sin\theta = \sin\abs{\frac{\phi_j - \phi_{j'}}{2}}\\
\implies
  \abs{\frac{\phi_i - \phi_{i'}}{2}} = \abs{\frac{\phi_j - \phi_{j'}}{2}} = \theta,
\end{multline}
with the sign determined by the relative direction of $\mathbf{p}$ and $\mathbf{q}$.
Since $\phi = \phi_q + \chi\frac{\pi}{2}$,
\begin{equation}
    \sgn(\phi_i - \phi_{i'}) = -\chi
\end{equation}
and similarly
since $\phi' = \phi_q + \chi'\frac{\pi}{2}$,
\begin{equation}
    \sgn(\phi_j - \phi_{j'}) =  \chi'.
\end{equation}
Thus,
\begin{equation}
  \frac{\phi_i - \phi_{i'}}{2} = -\chi \theta,\quad
 \frac{\phi_j - \phi_{j'}}{2} = \chi' \theta.
 \label{eq:anglediff}
\end{equation}
Now for the sums of angles
\begin{gather}
  \begin{multlined}[][\arraycolsep]
    \tan\phi = \frac{\sin\phi_i + \sin\phi_{i'}}{\cos\phi_i + \cos\phi_{i'}} = \tan\frac{\phi_i + \phi_{i'}}{2}\\
    \implies \phi = \frac{\phi_i + \phi_{i'}}{2}
\end{multlined}\\
\begin{multlined}[][\arraycolsep]
  \tan\phi' = \frac{\sin\phi_j + \sin\phi_{j'}}{\cos\phi_j + \cos\phi_{j'}} = \tan\frac{\phi_j + \phi_{j'}}{2}\\
  \implies \phi' = \frac{\phi_j + \phi_{j'}}{2}.
\end{multlined}
 \label{eq:anglesum}
\end{gather}

Combining \cref{eq:anglediff,eq:anglesum} we have
\begin{equation}
\begin{gathered}
 \phi_i = \phi - \chi\theta = \phi_q + \chi(\frac{\pi}{2} - \theta)\\
 \phi_{i'} = \phi + \chi\theta = \phi_q + \chi(\frac{\pi}{2} + \theta)\\
 \phi_j = \phi' + \chi'\theta = \phi_q + \chi'(\frac{\pi}{2} + \theta)\\
 \phi_{j'} = \phi' - \chi'\theta = \phi_q + \chi'(\frac{\pi}{2} - \theta).
\end{gathered}
\end{equation}
\begin{figure*}[tp]
  \centering
  \includegraphics[width=0.3\linewidth]{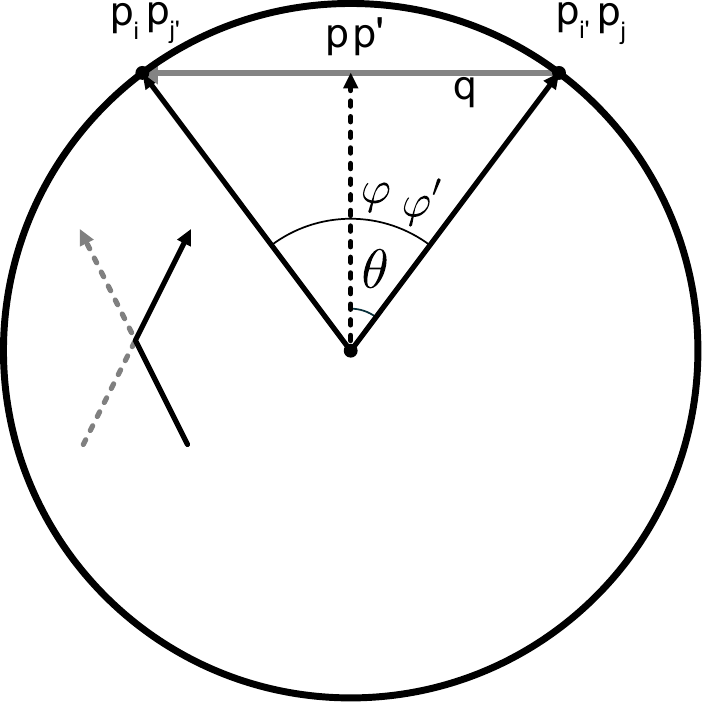}
  \qquad
  \includegraphics[width=0.3\linewidth]{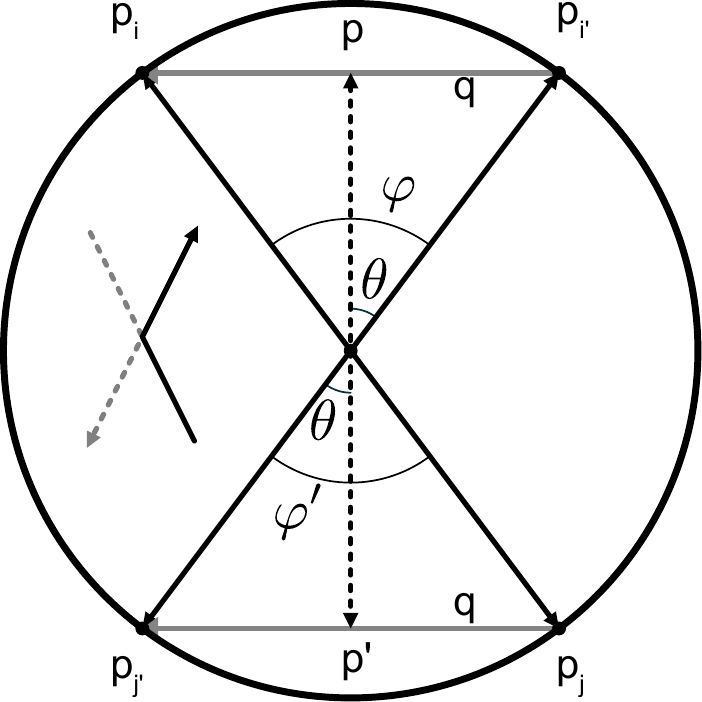}
  \caption{Position of scattered particles on the Fermi surface and associated angles, for collinear (left) and head-on (right) scattering processes. Note that the momenta $\mathbf{p}$ and $\mathbf{p}'$ do not lie on the Fermi surface.\label{fig:fsscatter}}
\end{figure*}

\begin{widetext}
We consider the two cases, $\chi = \chi'$ and $\chi = -\chi'$ corresponding to the left and right hand diagrams in \cref{fig:fsscatter}.
When $\chi = \chi'$,
\begin{equation}
 \phi = \phi' = \phi_q + \chi \pi/2
\end{equation}
and we have collinear scattering
\begin{equation}
  \begin{gathered}
  \phi_i = \phi_{j'} =  \phi_q + \chi(\frac{\pi}{2} -\theta),\\
  \phi_j = \phi_{i'} =  \phi_q + \chi(\frac{\pi}{2} + \theta).
\end{gathered}
\end{equation}
In the other case
\begin{equation}
 \phi = \phi_q + \chi \frac{\pi}{2} = \phi_q -\chi \frac{\pi}{2} + \pi  = \phi' + \pi
\end{equation}
which describes head on scattering
\begin{equation}
  \begin{gathered}
  \phi_i = \phi_{j} + \pi  =\phi_q + \chi(\frac{\pi}{2} -\theta),\\
  \phi_{i'} = \phi_{j'} + \pi = \phi_q + \chi(\frac{\pi}{2} + \theta).
\end{gathered}
\end{equation}
The corresponding contributions to the collision integral are
\begin{multline}
  I^{\mu;For.}_{m,-}[\eta^\mu]
  = -
  2\sum_{m'}\eta^\mu_{m'}
  \frac{\nu_F T^2}{v_F^2}\int d\Sigma
  \int_0^\frac{\pi}{2} \frac{d\theta}{\sin\theta\cos\theta}
W^\mu_{-}(0, \theta)
  \sum_{\chi}\oint \frac{d\phi_q}{2\pi}\\
  \times
    e^{-im(\phi_q + \chi\pi/2 - \chi\theta)}
   \left(
    e^{im'(\phi_q + \chi\pi/2 - \chi\theta)} -   e^{im'(\phi_q + \chi\pi/2 + \chi\theta)}\right)\\
  = -2
  \eta^\mu_{m}
  \frac{\nu_F T^2}{v_F^2}\int d\Sigma
  \int_0^\frac{\pi}{2} \frac{d\theta}{\sin\theta\cos\theta}
  W^\mu_{-}(0, \theta)
  \sum_{\chi}
   \left(
    1 -   e^{2im\chi\theta}\right)\\
  = -
  4\eta^\mu_{m}
  \frac{\nu_F T^2}{v_F^2}\int d\Sigma
  \int_0^\frac{\pi}{2} \frac{d\theta}{\sin\theta\cos\theta}
  W^\mu_{-}(0, \theta)
   \left(
    1 -   \cos2m\theta\right)
\end{multline}
and
\begin{multline}
  I^{\mu;H.O.}_{m,-}[\eta^\mu]
  = -
  2\sum_{m'}\eta^\mu_{m}
  \frac{\nu_F T^2}{v_F^2}\int d\Sigma
  \int_0^\frac{\pi}{2} \frac{d\theta}{\sin\theta\cos\theta}
  \sum_\chi
W^\mu_{-}(\pi, \theta)
e^{-im(\phi_q + \chi\pi/2 - \chi\theta)}\\
\times
\left(
e^{im'(\phi_q + \chi\pi/2 - \chi\theta)}-   e^{im'(\phi_q + \chi\pi/2 + \chi\theta )} -e^{im'(\phi_q + \chi\pi/2 - \chi\theta + \pi)} +   e^{im'(\phi_q + \chi\pi/2 + \chi\theta + \pi)}\right)\\
  = -2
  \eta^\mu_{m}
  \frac{\nu_F T^2}{v_F^2}\int d\Sigma
  \int_0^\frac{\pi}{2} \frac{d\theta}{\sin\theta\cos\theta}
W^\mu_{-}(\pi, \theta)
  \sum_{\chi}
\frac{1}{2} \left(1 - e^{im\pi}\right)
\left(
1 -   e^{2im \chi\theta }\right)\\
  = -4
  \eta^\mu_{m}
  \frac{\nu_F T^2}{v_F^2}\int d\Sigma
  \int_0^\frac{\pi}{2} \frac{d\theta}{\sin\theta\cos\theta}
W^\mu_{-}(\pi, \theta)
\frac{1}{2} \left(1 - e^{im\pi}\right)
\left(
1 -   \cos2m \theta \right)
\end{multline}
Combining the above two expressions, we can identify the contributions to the scattering rate at order $T^2$
\begin{equation}
 \frac{1}{\tau^\mu_{m}}  =
  \frac{8\nu_F T^2}{v_F^2}
  \int d\Sigma
  \int_0^\frac{\pi}{2} \frac{d\theta}{\sin\theta\cos\theta}
  \left(
   W^\mu_{-}(0, \theta) +  \frac{(1 - e^{im\pi})}{2}W^\mu_{-}(\pi, \theta)\right)
    \sin^2m\theta.
\end{equation}
\end{widetext}
Separating into even and odd $m$ terms 
\begin{multline}
 \frac{1}{\tau^\mu_{m, \text{even}}}  =
 \frac{8\nu_F T^2}{v_F^2} 
  \int d\Sigma
  \int_0^\frac{\pi}{2} \frac{d\theta}{\sin\theta\cos\theta}\sin^2m\theta\\
  \times
   \left(W^\mu_{-}(0, \theta)+W^\mu_{+}(\pi, \theta)\right)
    \sim T^2 \log m
  \label{eq:relax-even}
\end{multline}
and
\begin{multline}
 \frac{1}{\tau^\mu_{m,\text{odd}}}  =
 \frac{8\nu_F T^2}{v_F^2} 
  \int d\Sigma
  \int_0^\frac{\pi}{2} \frac{d\theta}{\sin\theta\cos\theta}\sin^2m\theta\\
  \times
  \left(
   W^\mu_{-}(0, \theta) +  W^\mu_{-}(\pi, \theta)\right)
  \label{eq:relax-odd}
\end{multline}
where we have restored the usual contribution~\cite{Ledwith2019} from the $+$ channel for the even modes.
Note that \cref{eq:relax-odd} is logarithmically divergent at $\theta \to \frac{\pi}{2}$, corresponding to $q=2p_f$, if $W(0, \pi/2) + W(\pi, \pi/2)$ is finite.
From \cref{fig:fsscatter}, one can see that then that the divergence is due to backscattering, $\mathbf{p}_i \to - \mathbf{p}_{i}$ and $\mathbf{p}_j \to - \mathbf{p}_j$.

These can be put into a more conventional form by writing in terms of the scattering angle $\theta_\text{sc} =2\theta$
\begin{multline}
 \frac{1}{\tau^\mu_{m, \text{even}}}  =
  \frac{4\nu_F T^2}{v_F^2}
  \int d\Sigma
  \int_0^\pi \frac{d\theta_\text{sc}}{\sin\theta_\text{sc}}(1 - \cos m\theta_\text{sc})\\
  \times
  \left( W^\mu_{-, \text{Collinear}}(\theta_\text{sc})
 +  W^\mu_{+, \text{Head On}}(\theta_\text{sc})
  \right)
\end{multline}
and
\begin{multline}
 \frac{1}{\tau^\mu_{m,\text{odd}}}  =
  \frac{4\nu_F T^2}{v_F^2}
  \int d\Sigma
  \int_0^\pi \frac{d\theta_\text{sc}}{\sin\theta_\text{sc}}(1-\cos m \theta_\text{sc})\\
  \times
  \left(
   W^\mu_{-, \text{Collinear}}(\theta_\text{sc}) +  W^\mu_{-, \text{Head On}}(\theta_\text{sc})\right)
   \label{eq:modd}
\end{multline}
where the subscript `sc' on the rates $W$ is just to show that we are now considering them as functions of $\theta_\text{sc}$.

The structure of the rates, as a sum of head-on and collinear terms, is a general feature of relaxation in 2D, as these are the only collision channels allowed by momentum and energy conservation for fermions on a circular Fermi surface (see e.g. Ref.~\cite{Laikhtman1992} or the supplement to Ref.~\cite{Alekseev2020}).
In the charge channel, the collinear term does not contribute to relaxation~\cite{Laikhtman1992,Ledwith2017,Ledwith2019} due to the indistinguishability of the scattering particles, but in general any collinear collision of particles with different values of a quantum number contributes to the angular relaxation of the associated channel, e.g. collinear collision of opposite spin Fermions contributes to the relaxation of the spin channel.
Similarly, the head on collision term vanishes at order $T^2$ for the charge channel~\cite{Ledwith2019}, but in general contributes for collisions with differing values of quantum number in the channel of interest.

There are two leading contributions to these integrands.
For odd terms, there is a logarithmic divergence at $\theta_\text{sc} \to \pi$.
For all terms there is also, potentially, a logarithmic contribution from forward scattering due to the long ranged Coulomb interaction.
Thus to leading log order the transport lifetimes can be approximated
\begin{equation}
    \frac{1}{\tau^\mu_{tr}} \equiv \frac{1}{\tau^\mu_{1}} \approx
    \frac{1}{\tau^\mu_{1,\text{Backscatter}}}
    + \frac{1}{\tau^\mu_{1,\text{Forward}}}.
\end{equation}
We evaluate these two contributions separately below.

\subsection{Back-scattering contribution}
\label{sec:backscatter}
To leading log order we can resolve the backscattering divergence by taking into account the cutoff on the integration region due to kinematic constraints on particle scattering.
To do so, we recall that the energy conservation delta-functions \cref{eq:deltaomega,eq:epsdiff,eq:epsdef} imposed the constraint
\begin{equation}
 \sqrt{1 + y} - \sqrt{1-y} = \frac{\omega}{\epsilon},
 \, \sqrt{1 - y'} - \sqrt{1+y'} = \frac{\omega}{\epsilon}
\end{equation}
where we have defined
\begin{equation}
 y= \frac{v^2 pq}{\epsilon^2}\cos(\phi - \phi_q)
 \label{eq:ydef}
\end{equation}
and as before
\begin{equation}
 \epsilon^2 = v^2p^2 + \Delta^2 + \frac{v^2q^2}{4}.
\end{equation}
Expanding to lowest order in $w=\omega/T$,
\begin{multline}
  \sqrt{1+y} - \sqrt{1-y}
  = \sum_{k=0}^\infty
 \begin{pmatrix}
   1/2\\
   k
 \end{pmatrix}
 \left(y^k -{(-y)}^k\right)\\
  = 2\sum_{k=0}^\infty
 \begin{pmatrix}
   1/2\\
   2k+1
 \end{pmatrix}
 y^{2k+1} = \frac{wT}{\mu + uT}\\
 = \frac{w}{1 + uT/\mu} T/\mu \approx w\frac{T}{\mu}.
\end{multline}
Substituting back in \cref{eq:ydef} we have, to order $T/\mu$,
\begin{multline}
  \frac{v^2p_F}{\mu^2} q\sqrt{1 - \frac{q^2}{4p_F^2}}\cos(\phi - \phi_q)\\
  =
  \frac{\mu^2 - \Delta^2}{\mu^2}\sin\theta_\text{sc}\cos(\phi - \phi_q)
  = w \frac{T}{\mu}
\end{multline}
Since $\abs{\cos} \leq 1$ we must then have
\begin{equation}
\sin \theta_\text{sc} \geq w \frac{T \mu}{\mu^2 - \Delta^2}.
\end{equation}
which acts as a constraint on the integration region.
From this we define the cutoff angle
\begin{equation}
 \sin\theta_c \approx \theta_c \approx  w \frac{T \mu}{\mu^2 - \Delta^2}.
\end{equation}
Noting that the dominant region for the measure $d\Sigma$ will come from
\begin{equation}
 w \approx 1 + O(T/\mu)
\end{equation}
we then approximate this as
\begin{equation}
\theta_c \approx  \frac{T \mu}{\mu^2 - \Delta^2}.
\end{equation}
The leading-log backscattering contribution to the transport time may then be obtained with the change of variables $x=\cos(\theta_\text{sc}/2)$ in \cref{eq:modd}
\begin{multline}
 \frac{1}{\tau^\mu_{1,\text{Backscatter}}}\approx 
  \int d\Sigma
  \int_{\theta_c/2}^1
  \frac{dx}{x}
 \\
 \times
  \frac{2\nu_F T^2}{v_F^2}
  \left(
   W^\mu_{-, \text{Collinear}}(\pi) +  W^\mu_{-, \text{Head On}}(\pi)\right)\\
  = \frac{2\nu_F T^2}{v_F^2}
  \int d\Sigma
  \ln\frac{\mu^2-\Delta^2}{T \mu}
 \\
 \times
  \left(
   W^\mu_{-, \text{Collinear}}(\pi) +  W^\mu_{-, \text{Head On}}(\pi)\right)
\end{multline}
Using the fact that
\begin{multline}
  T \ll \sqrt{\mu^2 - \Delta^2}\implies T \ll \mu \\
  \implies \ln\frac{\sqrt{\mu^2 - \Delta^2}}{T} \gg  \ln\frac{\sqrt{\mu^2 - \Delta^2}}{\mu},
\end{multline}
we have
\begin{multline}
 \frac{1}{\tau^\mu_{1,\text{Backscatter}}}\approx 
  \frac{4\nu_F T^2}{v_F^2}
  \int d\Sigma
  \ln\frac{\sqrt{\mu^2-\Delta^2}}{T}
 \\
 \times
  \left(
   W^\mu_{-, \text{Collinear}}(\pi) +  W^\mu_{-, \text{Head On}}(\pi)\right).
\end{multline}
To evaluate the integral over $d\Sigma$ we rewrite it as
\begin{multline}
 \int d\Sigma =
   \frac{1}{{(2\pi)}^2}
  \int du_i f_i (1-f_i) \\
  \times\int du_j du_{i'} du_{j'} \delta(u_i + u_j - u_{i'} -u_{j'})\\
  \times
  \frac{1}{f_i (1 -f_i)} f_i f_j (1 -f_{i'}) (1 - f_{j'})
\end{multline}
where
\begin{equation}
 f_i = \frac{1}{1 + \exp(u_i)}.
\end{equation}
Using the appendix of Ref.~\onlinecite{Ledwith2019} this can be simplified to
\begin{multline}
 \int d\Sigma=
  \frac{1}{{(2\pi)}^2}
  \int du_i f_i (1-f_i)
  \frac{u_i^2 + \pi^2}{2}\\
  =
  -\frac{1}{{(2\pi)}^2}
  \int du_i \frac{\partial f_i}{\partial u}
  \frac{u_i^2 + \pi^2}{2}
  = \frac{1}{6}
  \label{eq:sigma-int}
\end{multline}
giving
\begin{multline}
 \frac{1}{\tau^\mu_{1,\text{Backscatter}}}\approx 
  \frac{2\nu_F T^2}{3v_F^2}
  \ln\frac{\sqrt{\mu^2-\Delta^2}}{T}
 \\
 \times
  \left(
   W^\mu_{-, \text{Collinear}}(\pi) +  W^\mu_{-, \text{Head On}}(\pi)\right).
   \label{eq:tauBs}
\end{multline}

\subsection{Forward-scattering contribution}
\label{sec:forward}
In two dimensions, the phase space for collinear collisions is logarithmically divergent~\cite{DasSarma1985,Hwang2008,Muller2009,Muller2011}.
This being the case, the scattering rate for the unscreened Coulomb interaction, gives a divergent contribution to the transport scattering rate, despite the usual $1-\cos\theta_\text{sc}$ suppression of forward scattering contributions.
Physically this divergence is cut off by the screening wavevector.
In the case where $q_{TF} \gg 2 k_F$ the Coulomb interaction is shortranged and one need not consider the forward scattering contribution, as it is effectively suppressed by the transport form factor.
However, in the limit $q_{TF} \ll 2 k_F$ the region of the integral $\sin \theta_\text{sc} \sim q_{TF}/2 k_F$ gives
\begin{equation}
W^\mu_-(\theta \to 0) \propto \frac{1}{{(q + q_{TF})}^2} \propto \frac{1}{{(\frac{q_{TF}}{2k_F} + \sin\theta)}^2}
\end{equation}
leading to a logarithmic contribution proportional to $T^2\log q^{-1}_{TF}$.

One may see that the the Coulomb term enters all the distinguishable rates.
Its leading log behavior then provides the forward scattering contribution
\begin{multline}
 \frac{1}{\tau^\mu_{1,\text{Forward}}}  \approx
 2^6\frac{{(2\pi e^2)}^2}{\kappa^2}
 \frac{\nu_F T^2}{v_F^2}\int d\Sigma\\
 \times
  \int_{\frac{\theta_c}{2}}^{\frac{\pi}{2}-\frac{\theta_c}{2}} d\theta\frac{\tan\theta}{{(q_{TF} + 2k_F\sin\theta)}^2}
\end{multline}
where $\kappa$ is the background dielectric constant and the leading prefactor $2^6=8\times2\times2\times2$ comes from the prefactor of the relxation rate, the factor of $2$ for distinguishable particles, the sum over forward and head-on collisions, and finally the two different eigenvalue-changing channels.
Defining the effective fine-structure constant
 $\alpha = (e^2/\kappa v)$
we can rewrite
\begin{multline}
 \frac{1}{\tau^\mu_{1,\text{Forward}}}  \approx
 2^6\cdot 4\pi^2\alpha^2
 \frac{\nu_F T^2 v^2}{v_F^2}\int d\Sigma\\
 \times
  \int_{\frac{\theta_c}{2}}^{\frac{\pi}{2}-\frac{\theta_c}{2}} d\theta\frac{\tan\theta}{{(q_{TF} + 2k_F\sin\theta)}^2}
\end{multline}
Letting $x=\sin\theta$
\begin{multline}
 \frac{1}{\tau^\mu_{1,\text{Forward}}}  \approx
 2^6\cdot \pi^2\alpha^2
 \frac{\nu_F T^2 v^2}{v_F^2}\int d\Sigma\\
 \times
  \int_{\sin\frac{\theta_c}{2}}^{\cos\frac{\theta_c}{2}} \frac{dx}{1 -x^2}x\frac{1}{{(x + a)}^2}
\end{multline}
where we have defined $a=q_{TF}/2k_F \ll1$.
Extracting the logarithmic divergence due to the Coulomb potential we have
\begin{multline}
 \frac{1}{\tau^\mu_{1,\text{Forward}}}  \approx
 2^6\cdot \pi^2\alpha^2
  \frac{\nu_F T^2 v^2}{k_F^2 v_F^2}\int d\Sigma\\
  \times
  \int_{\frac{\theta_c}{2}}^{1} dx \frac{x}{{(x + a)}^2}.
\end{multline}
In principle, as this term is dominated by small angle scattering, dynamical screening effects may become relevant ~\cite{Alekseev2020}.
However, when $T \ll v q_{TF}$ the logarithmic divergence is cut off by the Thomas-Fermi wavevector before dynamic screening becomes relevant and we are justified in using the static screening approximation.
We henceforth work in this limit and subsequently neglect the cutoff angle $\theta_c$ for forward scattering as it does not contribute at leading log order.
Changing variables to $y=x+a$
\begin{multline}
 \frac{1}{\tau^\mu_{1,\text{Forward}}}  \approx
 2^6\cdot \pi^2\alpha^2
  \frac{\nu_F T^2 v^2}{k_F^2 v_F^2}\int d\Sigma
  \int_{a}^{1} dy \frac{y-a}{y^2}\\
  =
 2^6\cdot \pi^2\alpha^2
  \frac{\nu_F T^2 v^2}{k_F^2 v_F^2}\int d\Sigma
  \ln\frac{\sqrt{\mu^2 - \Delta^2}}{v q_{TF}}
\end{multline}
where we have used
\begin{equation}
 \sqrt{\mu^2 - \Delta^2} = v k_F .
\end{equation}
Performing the integral over $d\Sigma$, we then have
\begin{equation}
 \frac{1}{\tau^\mu_{1,\text{Forward}}}  \approx
 \frac{2^6\pi^2\alpha^2}{6} \frac{v^4}{v_F^2} \frac{\nu_F T^2}{v^2k_F^2}
  \ln\frac{\sqrt{\mu^2 - \Delta^2}}{vq_{TF}}
\end{equation}
Making use of
\begin{equation}
 v_F = \frac{v^2k_F}{\mu}, \qquad \nu_F = \frac{\mu}{2\pi v^2}
\end{equation}
we have
\begin{equation}
  \frac{v^4}{v_F^2}\nu_F = \frac{\mu^3}{2\pi v^2 k_F^2} = \frac{\mu}{2\pi} \frac{\mu^2}{\mu^2 - \Delta^2}
\end{equation}
and arrive at
\begin{equation}
 \frac{1}{\tau^\mu_{1,\text{Forward}}}  \approx
 \frac{2^4\pi \alpha^2}{3} \mu \frac{\mu^2 T^2}{{(\mu^2 - \Delta^2)}^2}
  \ln\frac{\sqrt{\mu^2 - \Delta^2}}{vq_{TF}}.
  \label{eq:tau-coulomb}
\end{equation}

\section{Evaluation of the backscattering probabilities \texorpdfstring{$W^\mu(\pi)$}{W}}
\label{sec:Wback}
In this section we approximate the probability for backscattering in each channel, appearing in \cref{eq:tauBs} for the non-relativistic case $\Delta \gtrsim E_F$ and the relativistic case $\Delta \ll E_F$.

As noted in \cref{eq:scattering-rates,eq:Born}, the scattering probabilities $W^\mu$ entering the collision integral can be expressed in terms of the interaction functions $U^\mu$, \cref{eq:gint}.
Using \cref{eq:int-functions-estimate}, we may estimate these rates in terms of the long-ranged part of the Coulomb interaction $V(q)$ and the short-ranged interactions constants $g_{00}, g_{\perp\perp}, g_{zz}, \tilde{g}_{00}$.

As noted above, the backscattering contribution will contain the collinear $\mathbf{p} = \mathbf{p}',\mathbf{q} = -2\mathbf{p}$ and head-on $\mathbf{p} = -\mathbf{p}',\mathbf{q} = 2\mathbf{p}$ terms, as depicted in \cref{fig:fsscatter}.
Looking at this figure, it is clear that that for $\theta = \pi$, the head-on and collinear terms should be equal.

Combining \cref{eq:Born,eq:int-functions-estimate,eq:coherence} we find the combined collinear and head-on contributions in each channel $W^\mu_\text{BS} =  W^\mu_{-, \text{Collinear}}(\pi) +  W^\mu_{-, \text{Head On}}(\pi)$ are
\begin{widetext}
\begin{equation}
  \begin{gathered}
  W^s_\text{BS} =
  8\abs*{
      g_{zz}
      + \frac{\Delta}{\mu}\tilde{g}_{00}
      + V_{2k_F}\frac{\Delta^2}{\mu^2}
   }^2\\
   W^{vz}_\text{BS} = W^{mz}_\text{BS}= 4\left(
\abs*{
      g_{zz}
      + \frac{\Delta}{\mu}\tilde{g}_{00}
     +  V_{2k_F}\frac{\Delta^2}{\mu^2}
    - 2g_{\perp\perp}\left(1 + \frac{\Delta^2}{\mu^2}\right)
    }^2
    +
   \abs*{
      g_{zz}
      + \frac{\Delta}{\mu}\tilde{g}_{00}
      + V_{2k_F}\frac{\Delta^2}{\mu^2}
   }^2
 \right)
 \\
 W^{v\parallel}_\text{BS} =  4\left(
      \abs*{g_{zz}
      + \frac{\Delta}{\mu}\tilde{g}_{00}
     +  V_{2k_F}\frac{\Delta^2}{\mu^2}
    - 2g_{\perp\perp}\left(1 + \frac{\Delta^2}{\mu^2}\right)}^2
    + \abs*{
      g_{zz}
      + \frac{\Delta}{\mu}\tilde{g}_{00}
      + V_{2k_F}\frac{\Delta^2}{\mu^2}
      - g_{\perp\perp}\left(1 + \frac{\Delta^2}{\mu^2}\right)
   }^2
   \right)
 \\
 W^{m\parallel}_\text{BS}=  4\left(
      \abs*{g_{zz}
      + \frac{\Delta}{\mu}\tilde{g}_{00}
     +  V_{2k_F}\frac{\Delta^2}{\mu^2}
    - 2g_{\perp\perp}\left(1 + \frac{\Delta^2}{\mu^2}\right)}^2
    + \abs*{
      g_{zz}
      + \frac{\Delta}{\mu}\tilde{g}_{00}
      + V_{2k_F}\frac{\Delta^2}{\mu^2}
      + g_{\perp\perp}\left(1 + \frac{\Delta^2}{\mu^2}\right)
   }^2
   \right)
   \end{gathered}
   \label{eq:backscatter-rates}
\end{equation}
\end{widetext}

\subsection{Non-relativistic limit}
For the non-relativistic limit $V(2k_F)$ is the dominant energy scale due \cref{eq:interaction-scale-hierarchy}, and we can approximate \cref{eq:backscatter-rates} as simply
\begin{equation}
W^\mu_{BS} \approx 8 |V(2 k_F)|^2 \frac{\Delta^4}{\mu^4}.
\end{equation}
Plugging this into \cref{eq:tauBs} and combining with \cref{eq:tau-coulomb} leads to \cref{eq:taudelta} of the main text.

\subsection{Relativistic limit}
In the relativistic limit, the long range Coulomb contributions are suppressed by the factor $\Delta^2/\mu^2$ and the short-ranged interaction constants become important.
From \cref{eq:interaction-harmonics,eq:interaction-scale-hierarchy} we see that the leading order contribution to the scattering rates will be due to $g_{\perp\perp}$ for all channels except the spin channel, where it is absent and $g_{zz}$ is the leading term. 
We may thus approximate \cref{eq:backscatter-rates} as
\begin{equation}
  \begin{gathered}
  W^s_\text{BS} =
  2 \times \abs{ g_{zz} }^2\\
   W^{vz}_\text{BS} = W^{mz}_\text{BS}= 2\times
8\abs{ g_{\perp\perp} }^2
 \\
 W^{v\parallel}_\text{BS} =W^{m\parallel}_\text{BS} =  2 \times 10
      \abs{
 g_{\perp\perp}
    }^2
   \end{gathered}
\end{equation}
Again, plugging this into \cref{eq:tauBs} and combining with \cref{eq:tau-coulomb} leads to \cref{eq:taunodelta} of the main text.
\end{document}